\documentclass[useAMS,usenatbib,onecolumn]{mn2e}

\usepackage{graphicx}
\usepackage{float}
\usepackage{rotating}
\usepackage{epsfig,floatflt}

\newcommand{\rvac}{\rho_{\rm vac}}
\newcommand{\lle}{\Lambda{\rm LE}}

\newcommand{\dd}{{\rm d}}

\newcommand{\rl}{r_{\Lambda}}

\newcommand{\pa}{\partial}
\newcommand{\vr}{\textbf{r}}

\newcommand{\rn}{\textbf{r}}

\newcommand{\lp}{\left(}
\newcommand{\rp}{\right)}

\newcommand{\dtr}{\dd^{3}{\rm r}}

\newcommand{\maw}{\mathcal{W}}
\newcommand{\mak}{\mathcal{K}}

\newcommand{\mai}{\mathcal{I}}

\newcommand{\maa}{\mathcal{A}}

\begin{document}

\title[Probing Dark Energy at Astrophysical Scales]{Astrophysical Configurations with Background Cosmology: \\ Probing Dark Energy at Astrophysical Scales}

\author[Balaguera-Antol\'{\i}nez, Mota, Nowakowski]{A. Balaguera-Antol\'{\i}nez$^1$\thanks{E-mail: abalan@mpe.mpg.de}, D. F. Mota$^{2}$\thanks{E-mail: d.mota@thphys.uni-heidelberg.de} and M. Nowakowski$^3$\thanks{E-mail: mnowakos@uniandes.edu.co}
\\
$^1$Max Planck Institute f\"ur Extraterrestrische Physik, D-85748, Garching, Germany
\\ $^2$Institute for Theoretical Physics, University of Heidelberg, 69120 Heidelberg, Germany
%
\\$^3$ Departamento de F\'{\i}sica, Universidad de los Andes, A.A. 4976, Bogot\'a, D.C., Colombia
}
\maketitle

\begin{abstract}
 We explore the effects of a positive cosmological constant on astrophysical and cosmological configurations 
described by a polytropic equation of state. We derive the conditions for equilibrium and stability of such 
configurations and consider some astrophysical examples where our analysis may be relevant. 
We show that in the presence of the cosmological constant the isothermal sphere is not a viable astrophysical model 
since the density in this model does not go asymptotically to zero. The cosmological constant implies that, for polytropic index smaller
than five, the central density has to exceed a certain  minimal value in terms of the vacuum density in order to
guarantee the existence of a finite size  object. We examine such configurations together with 
effects of $\Lambda$ in other exotic possibilities, such as neutrino and boson stars, and we compare our results to N-body simulations. 
The astrophysical properties and configurations found in this article are specific features resulting from the existence
of a dark energy component.   Hence,  if
found in nature would be an independent probe of a cosmological constant, complementary 
to other observations.
\end{abstract}

\pagerange{\pageref{firstpage}--\pageref{lastpage}} \pubyear{2007}
\label{firstpage}
%

\section{Introduction}
Some of the most relevant properties of the universe have been established through astronomical 
data associated to light curves of distant Supernova Ia \cite{riess},
 the temperature anisotropies in the cosmic microwave background radiation \cite{wmap3}  and the matter power spectrum of large scale structures \cite{tegmark}. 
Such observations give a strong evidence
that the geometry of the universe is flat and that our Universe is undertaking an
accelerated expansion at the present epoch. This acceleration 
is attributed to a dominant dark 
energy component, whose most popular candidate is the 
cosmological constant, $\Lambda$.

The present days dominance of dark energy make us wonder if this component 
may affect the formation and stability of large astrophysical structures, whose physics is basically Newtonian. 
This is in fact an old question put forward already by Einstein \cite{einstein} and pursued by many other authors \cite{noer,chernin2,manera,nunes,baryshev} 
and \cite{kagramanova,jetzer1}.
In general, the problem is rooted
in the question whether the expansion of the universe, which in
the Newtonian sense could be understood as a repulsive force, affects local astrophysical
properties of large objects. The answer is certainly affirmative if part of the terms responsible for the Universe expansion survives the
Newtonian limit of the Einstein equations. This is indeed the case of $\Lambda$ which is part of the Einstein tensor.
In fact, as explained in the main text below,
all the effects of the universe expansion can be taken into account, regardless of the model, by generalizing the Newtonian limit. 
This approach allow us to calculate the impact of a given cosmological model on astrophysical
structures. 

Although, there are several candidates to dark energy which have their own cosmological signature, e.g. 
\cite{Koivisto1, daly, wang2, koivisto, koivisto2, shaw2} and \cite{seo,brook,daly2,Koivisto3, shaw3}, 
 in this paper we will investigate the $\Lambda$CDM model only.
Such consideration is in fact not restrictive and our results will be common to most dark energy models. At 
astrophysical scales and within the Newtonian limit one does  not expect to find important differences among the different dark energy models.  This, however should not be interpreted as if 
$\Lambda$ has no effect at smaller astrophysical scales. In fact, the effects of a cosmological constant 
on the equilibrium and stability of astrophysical structures is not negligible, and  can be of relevance to 
describe features of astrophysical systems such as globular clusters, galaxy clusters or even galaxies \cite{cher,iorio,bala1,nowakowski1,cardoso}. 
Motivated by this,  
we investigate the effects of a dark energy component on the Newtonian limit of Einstein gravity and its consequences at astrophysical scales. 

In this article we investigate how 
the cosmological constant changes certain aspects of astrophysical hydrostatic
equilibrium. In particular we search for specific imprints which are unique to the existence of a dark energy fluid.
For instance,  the instability of previously viable astrophysical models when $\Lambda$ is included.
We explore such possibility using spherical configurations 
described by a polytropic equation of state (e.o.s) $p\sim \rho^{\gamma}$. 
  The polytropic equation of state 
derives its importance from its success and consistency, and it is widely used in determining the properties of gravitational structures ranging from stars 
\cite{chan} to galaxies \cite{binney}. 
It leads to an acceptable description of the  
behavior of astrophysical objects in accordance with 
observations and numerical simulations \cite{kennedy1,gruzinov,kaniadakis,sadeth,pinzon1,ruffet}. 
The description of such configurations can be verified  in the general relativistic framework \cite{herrera} 
and applications of these models to the dark energy  problem have in fact been explored \cite{mukhop}. 

The effect of a positive cosmological constant can be best visualized as a repulsive non local force
acting on the matter distribution.  It is clear that this extra force will result
into a minimum density (either central or average) which is possible for the
distribution to be in equilibrium. 
This minimum density is a crucial crossing point: below this value no matter can be in equilibrium, above this value
low density objects exist \cite{lahav91}. Both effects are novel features due to $\Lambda$.
We will demonstrate such inequalities, which are generalizations of
corresponding inequalities found in \cite{nowakowski1} and \cite{bala2}, for every polytropic index $n$.
However, the most drastic effect can be found in the
limiting case of the polytropic equation of state, 
i.e,  the isothermal sphere where the polytropic index $n$ goes to infinity. 
This case captures, as far as the effects of $\Lambda$ are
concerned, many features also for higher, but finite $n$. The model
of the isothermal sphere is often used to model galaxies and galactic clusters \cite{natarajan} 
and used in describing effects of gravitational lensing \cite{kawano,sereno,maccio}. 
Herein lies the importance of the model.
Regarding the isothermal sphere we will show that $\Lambda$ renders the model unacceptable on general grounds.
This essentially means that the model does not even have an appealing asymptotic behavior for large radii and
any attempt to definite a physically acceptable radius has its severe drawbacks.

The positive cosmological constant offers, however, yet another unique opportunity, namely the
possible existence of young low density virialized objects, understood as configurations that have reached virial equilibrium just at the vacuum dominated epoch (in contrast to the structures forming during the matter dominated era, where the criteria for virialization is roughly $\bar{\rho}\approx 200 \rho_{\rm crit}$). This low density hydrostatic/virialized objects
can be explained again due to $\Lambda$ which now partly plays the role of the outward pressure.

The applicability of fluid models, virial theorem and hydrostatic equilibrium to large astrophysical bodies has been 
discussed many times in the literature. For a small survey on this topic we refer the reader to \citep{jackson,bala5} where one can also
find the relevant references.

It is interesting to notice that  dark matter halos represent a constant density background
which, in the Newtonian limit, objects embedded in them feel the analog to a negative cosmological constant. The
equilibrium analysis for such configurations has been performed in \cite{Umemura, Horedt}.
A negative $\Lambda$ will just enhance the attractive gravity effect, whereas a positive one opposes this attraction.
As a result the case $\Lambda > 0$ reveals different physical concepts as discussed in this paper.

The article is organized as follows. In the next section we introduce the equations relevant
for astrophysical systems as a result of the weak field 
limit and the non-relativistic limit of Einstein field equations taking into account a cosmological constant. 
There we  derive such limit
taking into account the background expansion independently of the dark energy model.
In section 3 we derive the equations governing polytropic configurations, 
the equilibrium conditions and stability criteria.
In section 4 we describe the isothermal sphere and investigate
its applicability in the presence of $\Lambda$. 
In section 5 we explore some examples of astrophysical configurations 
where the cosmological constant may play a relevant role. 
In particular, we probe into low density objects, fermion (neutrino) stars and boson stars.
Finally we perform an important comparison between polytropic configurations with $\Lambda$ and 
parameterized density of Dark Matter Halos.
We end with conclusions. We use  units $G_N=c=1$ except in section 5.4 where we restore $G_N$ and use natural units $\hbar=c=1$.

\section{Local dynamics in the cosmology background}
The dynamics of the isotropic and homogeneous cosmological background is determined 
by the evolution of the (dimensionless) scale factor given through the Friedman-Robertson-Walker 
line element as solution of Einstein field equations,
\begin{equation}\label{fri}
\frac{\ddot{a}(t)}{a(t)}=-\frac{4}{3}\pi \left[\rho(t)+3p(t)\right],
\hspace{0.5cm}\left[\frac{\dot{a}(t)}{a(t)}\right]^{2}=H(t)^{2}=\frac{8}{3}\pi\rho(t)-\frac{k}{a^{2}(t)},
\end{equation}
corresponding to the Raychaudhury equation and Friedman equation, respectively. The total energy density $\rho$ 
is a contribution from a matter component - baryonic plus dark matter - ($\rho_{\rm mat}\sim a^{-3}$), 
radiation ($\rho_{\rm rad}\sim a^{-4}$) and a dark energy 
component ($\rho_{\rm x}\sim a^{-f(a)}$ with $p=\omega_{\rm x}\rho$). The function $f(a)$ is given as 
\begin{equation}\label{fa}
f(a) \equiv \frac{3}{\ln a}\int_{1}^{a}\frac{\omega_{\rm x}(a')+1}{a'}\dd a', 
\end{equation}
where the term $\omega_{\rm x}(a)$ represents the equation of state for the dark energy component. 
The case $\omega_{\rm x}=-1$ corresponds to the cosmological constant $\rho_{\rm x}=\rvac=\Lambda/8\pi$.
The effects of the background on virialized structures can be explored through the Newtonian 
limit of field equations from which one can derive a 
modified Poisson's equation (see e.g. \cite{noer,nowakowski2}). 
Recalling that pressure is also a source for gravity, the gravitational potential produced by an overdensity is given by 
\begin{equation}\label{o}
\nabla^{2}\Phi=4\pi (\rho_t+3P_t),
\end{equation}
where  $\rho_t=\delta\rho + \rho$  and $P_t=\delta P + P$. Where $\delta \rho$ is the local overdensity with respect to the background density $\rho$. 
Notice that equation (\ref{o}) reduces to the usual Poisson equation, $\nabla^{2}\Phi=4\pi \delta  \rho$, when non relativistic matter dominates the Universe, 
and $\delta \rho \gg \rho$. However, at present times, when  dark energy dominates, the pressure is non-negligible and $\delta P$ might even be non zero, such 
as in the case of quintessence models \cite{ml,wang,mota1}. In this work, however, we will focus in the case of an homogeneous dark energy component where $\delta\rho=\delta P=0$. 
With this in mind, one can then write the modified Poisson equation as
\begin{equation} \label{yyy0}
\nabla^{2}\Phi=4\pi \delta  \rho-3\frac{\ddot{a}(t)}{a(t)}, 
\end{equation}
Note that this equation allows one to probe local effects of different Dark Energy models through the term $\ddot{a}/a$ given in Eq.(\ref{fri}).

Since we will be investigating the configuration and stability of astrophysical objects nowadays, when dark energy dominates, 
it is more instructive to write the above
equations in terms of an effective vacuum density i.e.
\begin{equation}\label{pois}
\nabla^{2}\Phi=4\pi \delta  \rho-8\pi\rvac^{\rm eff}(a),
\end{equation}
where by using (\ref{fri}) $\rvac^{\rm eff}(a)$ has been defined as
\begin{equation}
\label{rhoeff}
\rvac^{\rm eff}(a)\equiv -\frac{1}{2}\left[\lp\frac{\Omega_{\rm cdm}}{\Omega_{\rm vac}}\rp a^{-3} 
+(1+3\omega_{\rm x})a^{-f(a)}\right]\rvac,
\end{equation} 
which reduces to $\rvac$ for $\omega_{\rm x}=-1$ and negligible contribution 
from the cold dark matter component with respect to the over-density $\delta \rho$. With $\Phi_{\rm grav}$ being the solution associated to the pure gravitational interaction, the full solution for the potential can be simply written as  
\begin{equation} \label{yyy1}
\Phi(r,a)=\Phi_{\rm grav}(r)-\frac{4}{3}\pi \rvac^{\rm eff}(a)r^{2},\hspace{1cm}\Phi_{\rm grav}(r)=-\int_{V'} \frac{\delta \rho(\rn')}{|\rn-\rn'|}\,\dtr',
\end{equation} 
which defines the Newton-Hooke space-time for a scale factor close to
the present time (vacuum dominated epoch), $\omega_{\rm x}=-1$ and $\Omega_{\rm cdm}\ll\Omega_{\rm vac}$ \cite{gibbons,aldro}. For a $\Lambda$CDM universe with $\Omega_{\rm cdm}=0.27$ and $\Omega_{\rm vac}=0.73$ we get $\rvac^{\rm
eff}(\rm today)=0.81\rvac$: that is, the positive density of matter which has an attractive effect
opposing the repulsive one of $\Lambda$ reduces effectively the
strength of the 'external force' in (\ref{pois}).  Note that,  although in the
text we will use the notation $\rho_{\rm vac}$ which would be vaild in the case of a Newton-Hook space time, it must be understood that
we can replace $\rvac$ by $\rvac^{\rm eff}(\rm today)=0.81\rvac$ for a $\Lambda$CDM background. 

Given the potential $\Phi(r,a)$, we can write the Euler's equation for a self-gravitating configuration as
$$\rho \frac{d \langle v_{i}\rangle}{d t}+ \partial_{i}p+ \rho \partial_{i} \Phi=0$$, where $\langle v_{i}\rangle$ is the (statistical) mean velocity and $\rho$ is the total energy density in the system. We can go beyond  Euler's equation and write down the tensor virial equation  which reduces to its 
scalar version for spherical configurations. The (scalar) virial equation with the background contribution reads as \cite{bala3,caimmi}
\begin{equation}
\label{virialeq}
\frac{\dd ^{2}\mai}{\dd t^{2}}=2T+\maw^{\rm grav}+3\Pi+\frac{8}{3}\pi \rvac^{\rm eff}(a)\mai-\int_{\pa V}p\lp\vec{r}\cdot \hat{n}\rp\,\dd A,
\end{equation} 
where $\maw^{\rm grav}$ is the gravitational potential energy defined by
\begin{equation}
\label{maw}
\maw^{\rm grav}=\frac{1}{2}\int_{V}\rho(\vr)\Phi_{\rm grav}(\vr) \dtr,
\end{equation}
and $T=\frac{1}{2}\int_{V}\rho\langle v^{2}\rangle \dtr$ is the contribution of ordered motions to the kinetic energy. 
Also, $\mai\equiv \int_{V} \rho r^{2}\dtr$ is the moment of inertia about the center of 
the configuration and $\Pi\equiv \int _{V}p\dtr$ is the trace of the dispersion tensor.
The full description of a self gravitating configuration is completed with an equation for mass conservation, energy conservation and an 
equation of state $p=p(\delta \rho,s)$. If we assume equilibrium 
via $\ddot{\mai}\approx 0$, we obtain the known virial theorem \cite{jackson,bala3}
\begin{equation}
\label{virial}
|\maw^{\rm grav}|=2T+3\Pi+\frac{8}{3}\pi \rvac^{\rm eff}(a)\mai,
\end{equation} 
where we have neglected the surface term in (\ref{virialeq}), which is
valid in the case when we define the boundary of the configuration
where $p=0$.  

With $\rvac^{\rm eff}$ given in (\ref{rhoeff}), one must
be aware that an equilibrium configuration is at the most a dynamical
one.  This is to say that the 'external repulsive force' in
(\ref{rhoeff}) is time dependent through the inclusion of the background expansion and so are the terms in (\ref{virialeq}). This  leads to a violation of energy
conservation, which also occurs in the virialization process \cite{wang3,ml, caimmi,mota1,shaw}). 
Traced over cosmological times,  this implies that if we insist on the second derivative of the
inertial tensor to be zero, then the internal properties of the object
like angular velocity or the internal mean velocity of the components
will change with time. 
Even if in the simplest case, one can assume that the
objects shape and its density remain constant. 
Hence, equilibrium here can be thought of as represented by long time averages in which case the second
derivative also vanishes, not because of constant volume and density,
but because of stability \cite{bala4,bala5}. 

The expressions
derived in the last section, especially (\ref{pois}) and
(\ref{virial}) can be used for testing dark energy models on
configurations in a dynamical state of equilibrium.  
However, in these cases,  one should point out that, in this approach there is no energy conservation within the overdensity:  dark energy flows in and out of the overdensity. 
Such feature is a consequence of the assumption that dark energy does not cluster at small scales (homogeneity of dark energy).  
This is in fact the most common assumption in the literature\cite{wang,chen,hore}, with a few exceptions investigated in \cite{wang3,ml, caimmi,mota1}.

In this paper, we
will concentrate on the possible effects of a background dominated by
a dark energy component represented by the cosmological constant at late times ($z\leq $1). It implies that the total density
involved in the definitions of the integral quantities appearing in
the virial equation can be approximated to $\rho\sim \delta \rho$. In
that case the Poisson equation reduces to the form
$\nabla^{2}\Phi=4\pi \delta \rho-8\pi \rvac$. As mentioned before, the
symbol $\rvac$ has no multiplicative factors in the case of a
Newton-Hooke space time, while for a $\Lambda$CDM model it must be
understood as $\rvac\to 0.8\rvac$. As the reader will see, the most
relevant quantities derived here come in forms of ratio of a
characterizing density and $\rvac$, and hence the extra factor
appearing in the $\Lambda$CDM can be re-introduced in the
characterizing density. For general
consideration of equilibrium in the spherical case see \cite{boehmer1}
and \cite{boehmer}, while the quasi spherical collapse with cosmological
constant has been discussed in \cite{debnath}.

\section{Polytropic configurations and the $\Lambda$-Lane-Emden equation}
We can determine the relevant features of astrophysical systems by
solving the dynamical equations describing a self gravitating
configuration (Euler's equation, Poisson's equation, continuity
equations).  In order to achieve this goal we must first know the
potential $\Phi$ to be able to calculate the gravitational potential
energy. To obtain $\Phi$, one must supply the density profile and
solve Poisson's equation. In certain cases, the potential is given and
we therefore can solve for the density profile in a simple way.  Here
we face the situation where no information on the potential (aside
from its boundary conditions) is available and we also do not have an
apriori information about the density profile (see for instance
\cite{binney} for related examples).  In order to determine both, the
potential and the density profile, a complete description of
astrophysical systems required. This means we need to know an equation
of state $p=p(\rho)$ (here we change notation and we call $\rho$ the
proper density of the system). The equation of state can take several
forms and the most widely used one is the so-called \emph{polytropic
equation of state}, expressed as
\begin{equation}
\label{poly}
p=\kappa \rho^{\gamma},\hspace{2cm}\gamma\equiv 1+\frac{1}{n},
\end{equation} 
where $\gamma$ is the polytropic index and $\kappa$ is a parameter
that depends on the polytropic index, central density, the mass and
the radius of the system. The exponent $\gamma$ is defined as
$\gamma=(c_{\rm p}-c)/(c_{\rm v}-c)$ and is associated with processes
with constant (non-zero) specific heat $c$. It reduces to the
adiabatic exponent if $c=0$. The polytropic equation of state was
introduced to model fully convective configurations.  From a
statistical point of view, Eq. (\ref{poly}) represents a
collisionless system whose distribution function can be written in
the form $f=f(\tilde{E})\sim \tilde{E}^{n-3/2}$, with $\tilde{E}\equiv
\phi-(1/2)mv^{2}$ being the relative energy and $\phi(r)\equiv
\Phi_{0}-\Phi(r)$ being the relative potential (where $\Phi_{0}$ is a
constant chosen such that $\phi(r=R)=0$)\cite{binney}.  In
astrophysical contexts, the polytropic equation of state is widely
used to describe astrophysical systems such as the sun, compact
objects, galaxies and galaxy clusters \cite{chan,binney,kennedy1}.
We now derive the well known Lane-Emden equation. We start from Poisson equation and Euler equation for spherically symmetric configurations, written as 
\begin{equation}
\label{poly+pois}
\frac{\dd ^{2}\Phi}{\dd r^{2}}+\frac{2}{r}\frac{\dd \Phi}{\dd r} =4\pi
\rho-8\pi \rvac, \hspace{1cm}\frac{\dd p}{\dd r}=-\rho\frac{\dd
\Phi}{\dd r}.
\end{equation} 
\begin{figure}
\begin{center}
\includegraphics[angle=0,width=12cm]{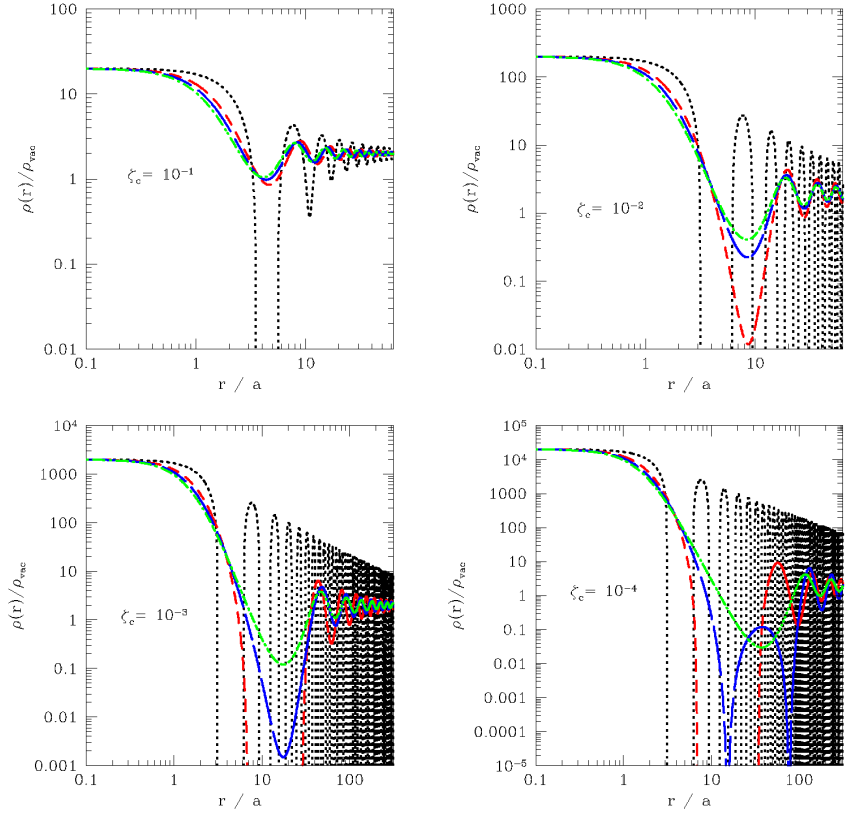}
\caption[]{{Solutions of $\Lambda$LE equation for different $\zeta_{c}$ and different index $n$ ranging from $n=3$ (red, short-dashed line), $n=4$ (blue,long-dashed line) and $n=5$ (green,dot short-dashed line).}}
\label{lanea}
\end{center}
\end{figure}
This set of equations together with Eq. (\ref{poly}) can be integrated
in order to solve for the density in terms of the potential as
\begin{equation}
\label{new1}
\rho(r)=\rho_{\rm c}\left[1-\lp\frac{\gamma-1}{\kappa
\gamma}\rp\rho_{\rm
c}^{1-\gamma}(\Phi(r)-\Phi(0))\right]^{\frac{1}{\gamma-1}},
\end{equation} 
where $\rho_c$ is the central
density.  In view of eq. (\ref{new1}) in conjunction with eq. (\ref{yyy1}) it
is clear that $\rvac$ will have the effect to increase the value of
$\rho(r)$. Therefore the boundary of the configuration will be located
in a greater $R$ as compared to the case $\rvac=0$. In order to determine the behavior of the density profile, we again combine Eq(\ref{poly+pois}) and Eq.(\ref{poly})  in order to eliminate the potential $\Phi(r)$. We obtain
\begin{equation}
\label{arbigeo2}
\frac{1}{n}\lp\frac{\nabla \rho}{\rho}\rp^{2}+\nabla^{2}\ln
\rho=-\frac{4\pi n\rho^{1-\frac{1}{n}}}{\kappa (n+1)}\lp1-\zeta \rp,
\end{equation} 
where we defined the function
\begin{equation} \label{yyy3}
\zeta=\zeta(r)\equiv 2\lp \frac{\rvac}{\rho(r)}\rp.
\end{equation} 
We can rewrite Eq (\ref{arbigeo2}) by introducing the variable $\psi$ defined by $\rho=\rho_{\rm c}\psi^{n}$, where $\rho_{\rm c}$ is the central density.
We also introduce the variable $\xi=r/a$ where
\begin{equation}
\label{a}
a\equiv \sqrt{\frac{\kappa (n+1)}{4\pi \rho_{\rm c}^{1-\frac{1}{n}}} }
\end{equation} 
is a length scale. Eq.(\ref{arbigeo2}) is finally written as \cite{bala2,chan}
\begin{equation}
\label{le}
\frac{1}{\xi^{2}}\frac{\dd }{\dd \xi}\lp \xi^{2}\frac{\dd \psi}{\dd
\xi}\rp=\zeta_{\rm c}-\psi^{n},
\hspace{0.8cm} \zeta_{\rm c} \equiv 2\lp\frac{\rvac}{\rho_{\rm c}}\rp.
\end{equation} 
This is the $\Lambda$-Lane-Emden equation ($\lle$). Note that for
constant density, we recover $\rho=2\rvac$ as the first non trivial solution of
$\lle$ equation. This is consistent with the results from virial
theorem for constant density spherical objects which tell us that
$\rho \ge 2\rho_{\rm vac}$ \cite{nowakowski1}.  Note that using Eq. (\ref{new1}) we can write the solution $\psi(\xi)$
with the explicit contribution of $\rvac$ as
\begin{equation}
\label{new2}
\psi\lp \xi=r/a\rp=1-\lp4\pi a^{2}\rho_{\rm c}\rp^{-1}(\Phi_{\rm grav}(r)-\Phi_{\rm grav}(0))+6\zeta_{\rm c}\xi^{2},
\end{equation} 
so that for a given $r$ smaller than the radius we will obtain
\begin{equation} \label{xxx6}
\psi\lp r/a\rp>\psi\lp r/a\rp_{\Lambda=0},
\end{equation} 
as already pointed out before.  Then the differential equation (\ref{le})
must be solved with the initial conditions $\psi(0)=1, \,\,\,
\psi'(0)=0$, satisfied by (\ref{new2}). Numerical solutions were
obtained for the first time in \cite{bala3}. 
The solutions presented
in Fig. \ref{lanea} are given in terms of the ratio $\rho/\rvac$ for
$n=3, 4$ and $n=5$. This choice of variables are useful also since
$\rvac$ sets a fundamental scale of density (the choice
$\rho_{0}=\rvac$ will be explored for the isothermal sphere, where
figure \ref{lanea} will be helpful for discussions).  The radius of a
polytropic configuration is determined as the value of $r$ when the
density of matter with the e.o.s (\ref{poly}) vanishes.  This happens
at a radius located at
\begin{equation} \label{xxx8}
R=a\xi_{1}\,\,\, {\rm such\,\, that} \,\,\, \psi(\xi_{1})=0.
\end{equation}
Note that equation (\ref{le}) yields a transcendental equation to
determine $\xi_{1}$.
Also one notes from Fig. \ref{lanea} that not all values of $\zeta_{\rm
c}$ yield allowed configurations in the sense that we cannot find a
value of $\xi_{1}$ such that $\psi(\xi_{1})=0$. This might not be
surprising since for $n=5$ and $\Lambda=0$ we find the situation where
the asymptotic behavior is $\rho \to 0$ as $\xi\to \infty$ (we
consider this still as an acceptable behavior). There is, however, one
crucial difference when we switch on a non-zero $\Lambda$.  For
$\Lambda\neq 0$ not only we cannot reach a definite radius but the
derivative of the density changes sign and hence becomes non-physical.
The situation for the cases $n \ge 5$ is somewhat similar to the
extreme case of $n \to \infty$ (isothermal sphere). Clearly, these features are responsible of the last term in Eq.(\ref{new2}), which for high values of $\zeta$ may become dominant over the remaining (gravitational) terms. We will discuss
this case in section four where we will attempt another definition of
a finite radius with the constraint $\psi(\xi)'<0$.  For now it
is sufficient to mention that, as expected, the radius of the allowed
configurations are larger than the corresponding radius when
$\Lambda=0$.

\subsection{Equilibrium and stability for polytropes}
In this section we will derive the equilibrium conditions for
polytropic configurations in the presence of a positive cosmological
constant. We will use the results of last section in order to write
down the virial theorem.  The total mass of the configuration can be
determined as usual with $M=\int \rho \dtr$ together with
Eq.(\ref{le}). One then has a relation between the mass, the radius and the central
density:
\begin{equation}
\label{radius3}
R=M^{1/3}\rho_{\rm c}^{-1/3}f_{0}(\zeta_{\rm c};n)=(Mr_{\Lambda}^{2})^{1/3}(4\pi )^{1/3}\zeta_{c}^{1/3}f_{0}(\zeta_{\rm c};n)
\end{equation}
where
\begin{equation}
f_{0}(\zeta_{\rm c};n)
\equiv
\lp\frac{\xi_{1}^{3}}{4\pi}\rp^{\frac{1}{3}}\lp\int_{0}^{\xi_{1}}\xi^{2}\psi^{n}(\xi)\dd\xi\rp^{-\frac{1}{3}},
\end{equation}
Note that we have introduced the cosmological constant in the equation for the radius, leading to the appearance of the astrophysical length scale $\lp Mr_{\Lambda}^{2}\rp^{1/3}$ (with $r_{\Lambda}=\Lambda^{-1/2}=(8\pi \rho_{\rm vac})^{-1/2}=2.4\times 10^{3}(\Omega_{\rm vac}h^{2})^{-1/2}$Mpc $\approx 4.14\times 10^{3}$ Mpc for the concordance values $\Omega_{\rm vac}=0.7$ and $h=0.7$). This scale has been already found in the
context of Schwarzschild - de Sitter metric where it is the maximum
allowed radius for bound orbits. At the same time it is the scale of
the maximum radius for a self gravitating spherical and homogeneous configuration in the presence of a positive $\Lambda$ \cite{bala3}. This also let us relate the mean density of the configuration with its central density and/or cosmological parameters as $\bar{\rho}=(3/4\pi f_{0}^{3}) \rho_{\rm c}=(3/2\pi \zeta_{\rm c}f_{0}^{3})\rho_{\rm vac}=(3\Omega_{\rm vac}/2\pi \zeta_{\rm c}f_{0}^{3})\rho_{\rm crit}$. 

Similarly we can determine the other relevant quantities appearing in
the equations for the energy and the scalar virial theorem
(\ref{virial}). For the traces of the moment of inertia tensor and the
dispersion tensor $\Pi$ we can write
\begin{equation}
\label{piner}
\mai=MR^{2}f_{1},\hspace{1cm}\Pi=\kappa \rho_{\rm
c}^\frac{1}{n}Mf_{2},
\end{equation}
where the functions $f_{1,2}$ have been defined as
\begin{equation}
\label{pintener}
f_{1}(\zeta_{\rm c};n)\equiv
\frac{\int_{0}^{\xi_{1}}\xi^{4}\psi^{n}\dd\xi}{\xi_{1}^{2}\int_{0}^{\xi_{1}}\xi^{2}\psi^{n}\dd\xi},
\hspace{1cm}f_{2}(\zeta_{\rm c};n)\equiv
\frac{\int_{0}^{\xi_{1}}\xi^{2}\psi^{n+1}\dd\xi}{\int_{0}^{\xi_{1}}\xi^{2}\psi^{n}\dd\xi},
\end{equation}
using (\ref{radius3}).  These functions are numerically determined
in the sequence $\psi(\xi;\zeta_{\rm c})\to \xi_{1}(\zeta_{\rm c})\to
f_{i}(\zeta_{\rm c})$, such that for a given mass we obtain the radius as $R=a(M;\xi_{1})\xi_{1}$.

\begin{figure}
\begin{center}
\includegraphics[angle=0,width=9cm]{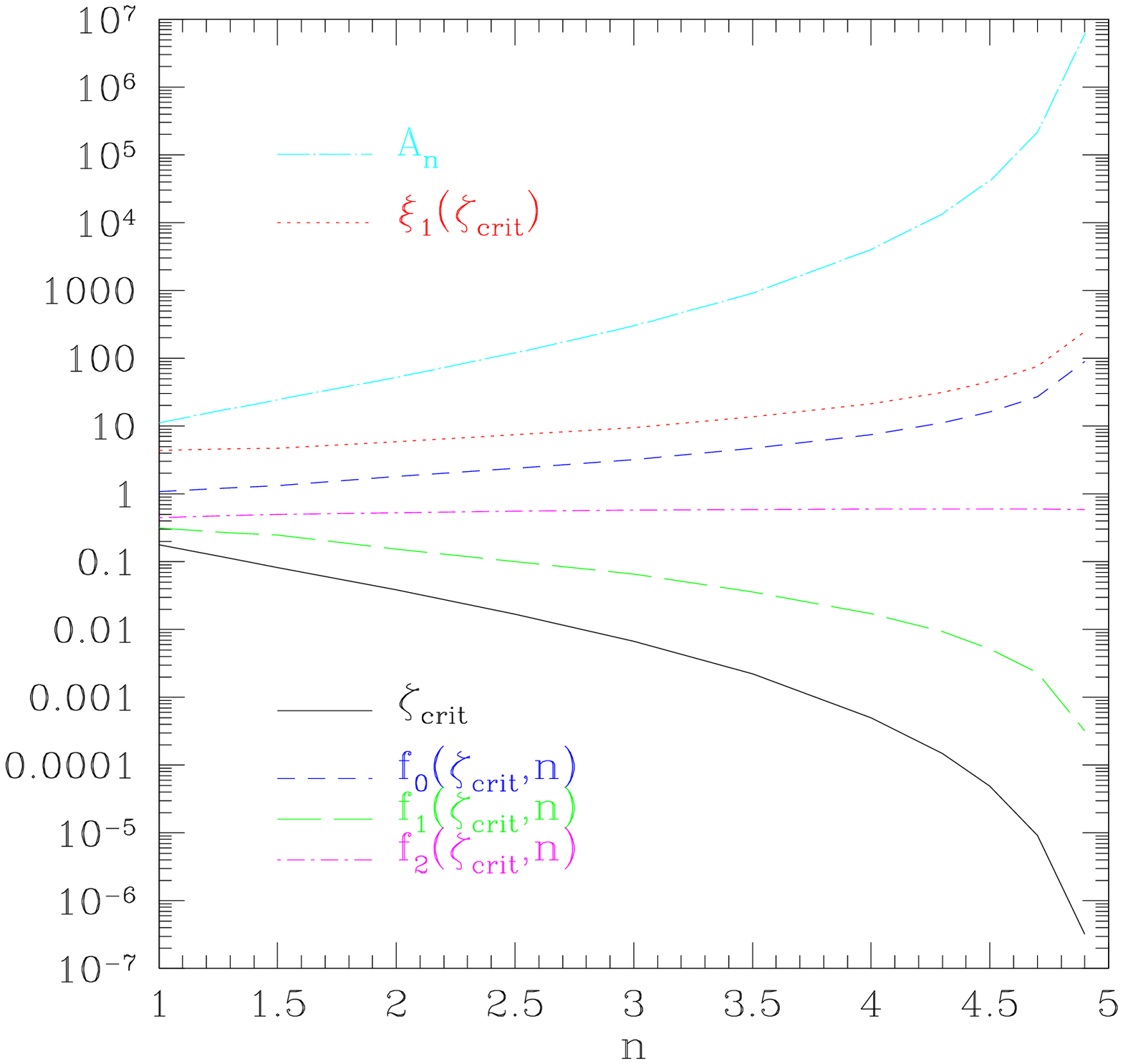}
\caption[]{{ The values of $\zeta_{\rm crit}$, $\xi(\zeta_{\rm crit})$, the functions $f_{i}(\zeta_{\rm c};n)$, and $\maa_{n}=2\zeta_{\rm crit}^{-1}$. Equilibrium configurations are reached for $\rho_{\rm c}>\maa
_{n}\rvac$ (for a $\Lambda$CDM cosmoolgy one has to rewrite $\maa_{n}\to 0.81\maa_{n}$).}}
\label{tablelane2}
\end{center}
\end{figure}

Let us consider the virial theorem (\ref{virial}) for a polytropic
configuration. The gravitational potential energy $\maw^{\rm grav}$
can be obtained following the same arguments shown in \cite{chan}.
The method consist in integrating Euler's equation and solve for
$\Phi_{\rm grav}$, then using  Eq.(\ref{maw}) one obtains $\maw^{\rm grav}$. 
The final result is written as
\begin{equation} 
\label{poliwc}
\maw^{\rm grav}=-\frac{M^{2}}{2R}-\frac{1}{2}(n+1)\Pi+\frac{2}{3}\pi\rvac\lp
\mai-MR^{2}\rp,
\end{equation}
To show the behavior of $\maw^{\rm grav}$ with respect to the index
$n$, we can solve the the virial theorem (\ref{virial}) for $\Pi$ and
replace it in Eq.(\ref{poliwc}). We obtain
\begin{equation} 
\label{poliwd}
\maw^{\rm
grav}=-\frac{3}{5-n}\left[1-\frac{\rvac}{\bar{\rho}}\lp\frac{1}{3}(5+2n)f_{1}-1\rp
\right]\frac{M^{2}}{R}.
\end{equation}
This expression shows the typical behavior of a $n=5$ polytrope (even
if $\rvac \neq0$): the configuration has an infinite potential energy,
due to the fact that the matter is distributed in a infinite volume.
The energy of the configuration in terms of the polytropic index can
be easily obtained by using (\ref{virial}), (\ref{poliwc}) and
(\ref{poliwd})
\begin{equation} 
\label{poliwen}
E=\maw^{\rm grav}+\frac{8}{3}\pi \rvac\mai
+n\Pi=-\lp\frac{3-n}{5-n}\rp\frac{M^{2}}{R}\left[1-
\frac{\rvac}{\bar{\rho}} \lp 5f_{1}^{(n)}-1\rp\right].
\end{equation}
One is tempted to use $E<0$ as the condition to be fulfilled for a
gravitationally bounded system.  For $\rvac=0$ we recover the condition
$\gamma>4/3$ ($n<3$) for gravitationally bounded configurations in
equilibrium. On the other hand, for $\rvac \neq 0$ this condition
might not be completely true due to the following reasoning:
The two-body effective potential in the presence of a positive
cosmological constant does not go asymptotically to zero for large
distances, which is to say that $E<0$ is not stringent enough to
guarantee a bound system.  Therefore, we rather rely on the numerical
solutions from which, for every $n$, we infer the value of ${\cal
A}_n$ such that
\begin{equation} \label{An}
\rho_c \ge {\cal A}_n \rho_{\rm vac}.
\end{equation}
This gives us the lowest possible central density in terms of
$\rho_{\rm vac}$. The behavior of the $\zeta_{\rm crit}$, the functions $f_{i}$, the solution $\xi_{1}(\zeta_{c}=\zeta_{\rm crit},n)$ and the values of ${\cal A}_n$ are shown in 
fig \ref{tablelane2}.  Note that this inequality can be understood as a
generalization of the equilibrium condition $\varrho>\maa \rvac$,
which, when applied for a spherical homogeneous configurations yields
$\maa=2$ with $\varrho=\rho=$ constant \cite{bala3,bala2}.

Note that, at $n=5$ the radius of the configuration becomes undefined, as well
as the energy. A $n=5$ polytrope is highly concentrated at the center
\cite{chan}. No criteria can be written since even for $\rvac=0$
there is not a finite radius.  But it is this high concentration at
the center and a smooth asymptotic behavior which makes this case
still a viable phenomenological model if $\Lambda$ is zero. On the
contrary for non-zero, positive $\Lambda$ the solutions start
oscillating around $2\rho_{\rm vac}$ which makes the definition of the
radius more problematic.  For $n\to \infty$ the polytropic e.o.s
describes an ideal gas (isothermal sphere).  Since in this limit the
expressions derived before are not well definite, this case will be
explored in more detail in the next section.  In spite of the
mathematical differences, the isothermal sphere bears many
similarities to the cases $n \ge 5$ and our conclusions regarding the
definition of a radius in the $n \to \infty$ case equally apply to
finite $n$ bigger than $5$.
\begin{figure}
\begin{center}
\includegraphics[angle=0,width=12cm]{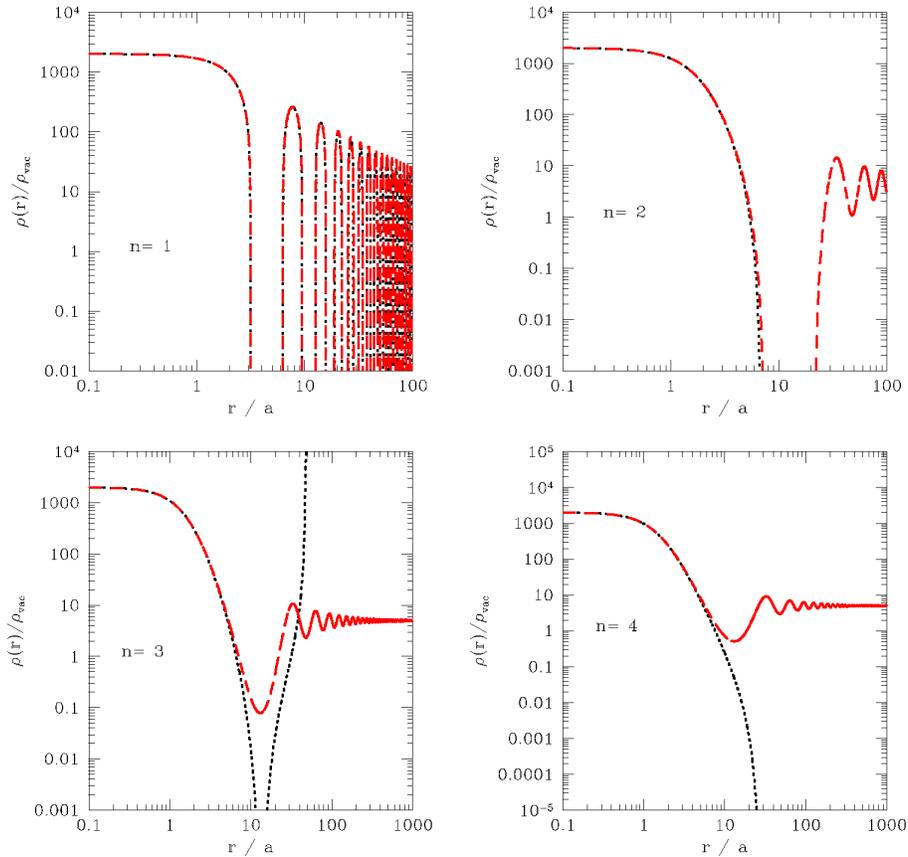}
\caption[]{{Density of a polytropic configuration for
 $\zeta_{\rm c}=10^{-3}$ and different polytropic index in a
modified Newton-Hook space time with a dark energy equation of state
$\omega_{\rm x}=-2/3$ (black,dots) and  $\omega_{\rm x}=-2$ (red,short dash). Compare with Fig.\ref{lanea}}}
\label{laneph}
\end{center}
\end{figure}

\subsection{Effects with generalized dark energy equation of state}
In the last section we have explored the effects of a dark
energy-dominated background with the equation of state
$\omega_{\rm x}=-1$. Other dark energy models are often used with
$\omega_{\rm x}=-1/3$ and $\omega_{\rm x}=-2/3$ or even $\omega_{\rm x}<-1$, in the so-called phantom regime, or even a time dependent dark energy model (quintessence). A simple generalization to such
models can be easily done by making the following replacement in our equations
\begin{equation}
\zeta\to \zeta(a)_{\rm eff}\equiv -\frac{1}{2}\zeta\eta(a) a^{-f(a)},
\end{equation}
where $\eta(a)=1+3\omega_{\rm x}(a)$, and where the function $f(a)$ is
defined in Eq.(\ref{fa}). Note that for this generalization to be  coherent with the derivation of
$\Lambda$LE in (\ref{le}), one needs to consider that the equation of state is close to $-1$, so that there is almost no time-dependence, 
and the energy density for dark energy is almost constant. This is indeed the case for most popular candidates of dark energy specially at low redshifts $z<1$.
Also, notice once again, that we are still assuming an homogeneous dark energy component which flows freely to and from the overdensity. 
Hence, violating energy conservation inside it.
Clearly, other models of dark energy will
posses dynamical properties that the cosmological constant does not
have. For instance, we could allow some fraction of dark energy to
take part in the collapse and virialization \cite{ml,caimmi,mota1},
which would lead to the presence of self and cross interaction terms
for dark energy and the (polytropic like) matter in Euler equation,
which at the end modifies the Lane Emden equation. With this
simplistic approach, we see that the effects with a general equation
of state are smaller than those associated to the cosmological
constant. In particular, the equation of state $\omega_{\rm x}=-1/3$
displays a null effects since it implies
$\eta=0$ (note that this equation of state can also resemble the curvature term in evolution equation for the background).
On the other hand, phantom models of dark energy, which are associated to equations of state
$\omega_{\rm x}<-1$\cite{cald,nojiri}, have quite a strong effect.
In Fig. \ref{laneph} we show numerical solutions of Lane-Emden
equations for a background dominated with dark energy with $\omega_{\rm x}=-2/3$ and a phantom dark energy with
$\omega_{\rm x}=-2$, with $\zeta=10^{-3}$. These curves are to be compared with those at fig.\ref{lanea}. Clearly equations of state with $\omega_{\rm
x}<-1$ will generate larger radius than the case described in the main
text. Furthermore, the asymptotic behavior of the ratio between the
density and $\rvac$ is $\rho/\rvac \to |\eta|$.

\subsection{Stability criteria with cosmological constant}
Stability criteria for polytropic configurations can be derived from the virial equation.  Using equations (\ref{piner}),
(\ref{pintener}) and (\ref{poliwc}) we can write the virial theorem  (\ref{virial}) in terms of the radius $R$ and the mass $M$: 
\begin{equation} 
\label{vpol}
-\frac{M^{2}}{2R}+\frac{1}{2}(5-n)\kappa \rho_{\rm
 c}^{\frac{1}{n}}Mf_{2}+\frac{2}{3}\pi \rvac MR^{2}\lp5f_{1}-1\rp=0,
\end{equation}
where we have assumed that the only contribution to the kinetic energy
comes from the pressure in the form of $\mak=\frac{3}{2}\Pi$. Note that for $\rvac=0$ and finite mass, one obtains $R\propto (5-n)^{-1}$
while for $\rvac\neq 0$ we would obtain a cubic equation for the
radius. Instead of solving for the virial radius, we solve for the mass as a
function of central density with the help of Eq (\ref{radius3}). We
have
\begin{equation} 
\label{fmas2aa}
M = \mathcal{G} \rho_{\rm c}^{\frac{3-n}{2n}},
\hspace{0.8cm}\mathcal{G} =\mathcal{G}(\zeta_{\rm c};n)\equiv
\left[\frac{\kappa f_{0}f_{2}(5-n)}{1-\frac{2}{3}\pi \zeta_{\rm
c}f_{0}^{3}(5f_{1}-1)}\right]^{\frac{3}{2}}.
\end{equation}
The explicit dependence of the mass with respect to the central
density splits into two parts: on one hand it has the same form as
the usual case with $\Lambda=0$, that is, $\rho_c^{(3-n)/2n}$; on the
other hand the function $\mathcal{G}$ has a complicated dependence on
the central density because of the term $\zeta_{\rm c}$.  With the
help of (\ref{radius3}) and (\ref{fmas2aa}) we can write a mass-radius
relation and the radius-central density relation
\begin{equation} 
\label{fmas2aaa}
M=\lp
\mathcal{G}^{\frac{2}{3}\lp\frac{n}{n-1}\rp}f_{0}^{\frac{3-n}{n-1}}\rp
R^{\frac{3-n}{1-n}},
\hspace{0.8cm}R=\lp \mathcal{G}^{\frac{1}{3}}f_{0}\rp\rho_{\rm
c}^{\frac{1}{2}\lp\frac{1-n}{n}\rp}.
\end{equation}
Following the stability theorem (see for instance in
\cite{weinberg}), the stability criteria can be determined from the
variations of the mass in equilibrium with respect to the central
density. We derive from Eq. (\ref{fmas2aaa}):
\begin{equation} 
\label{slope}
\frac{\pa M}{\pa \rho_{\rm
c}}=\left[\frac{3}{2}\lp\gamma-\frac{4}{3}\rp\rho_{c}^{-1}\mathcal{G}+\frac{\pa
\mathcal{G}}{\pa \rho_{\rm c}}\right]\rho_{\rm
c}^{\frac{3}{2}(\gamma-\frac{4}{3})}.
\end{equation}
Stability (instability) stands for $\pa M/\pa \rho_{\rm c}>0$ ($\pa
M/\pa \rho_{\rm c}<0$). This yields a critical value of the polytropic
exponent $\gamma_{\rm crit}$ when $\pa M/\pa \rho_{\rm c}=0$ given by
\begin{equation} 
\label{slope2}
\gamma_{\rm crit}=\gamma_{\rm crit}(\zeta_{c})\equiv
\frac{4}{3}+\frac{2}{3}\frac{\pa \ln \mathcal{G}}{\pa \ln \rho_{\rm
c}},
\end{equation}
in the sense that polytropic configurations are stable under small
radial perturbations if $\gamma>\gamma_{\rm crit}$.  It is clear that
the second term in (\ref{slope2}) also depends on the polytropic index
and therefore this equation is essentially a transcendental expression
for $\gamma_{\rm crit}$.  

It is worth mentioning that by including the
corrections due to general relativity, the critical value for
$\gamma_{\rm crit}$ is also modified as $\gamma_{\rm
crit}=(4/3)+R_{\rm s}/R$ \cite{shapiro} and hence for compact objects
the correction to the critical polytropic index is stronger from the
effects of general relativity than from the effects of the
background. This is as we would expect it. Stability of relativistic
configurations with non-zero cosmological constant has been explored
in \cite{boehmer, boehmer1}.

Going back to equation (\ref{fmas2aa}), we can write the mass of the
configuration as
$M=\alpha_{M}M(0)$,
where $M(0)$ is the mass when $\Lambda=0$ and
$\alpha_{M}=\alpha_{M}(\zeta_{\rm c},n)$ is the enhancement factor.
Both quantities can be calculated to give
\begin{equation} 
M(0)\equiv
\lp\kappa(5-n)f_{0}^{(n)}f_{2}^{(n)}\rp^{\frac{3}{2}}\rho_{\rm
c}^{\frac{3-n}{2n}},\hspace{0.8cm} \alpha_{M}\equiv \left[
\frac{f_{0}f_{2}}{f_{0}^{(n)}f_{2}^{(n)}\lp1-\frac{2}{3}\pi \zeta_{\rm
c}f_{0}^{3}(5f_{1}-1)\rp} \right]^{\frac{3}{2}},
\end{equation}
where $f_{i}^{(n)}\equiv f_{i}(\zeta_{\rm c}=0,n)$ are numerical
factors (tabulated in table 1) that can be determined in a
straightforward way.  Similarly, by using Eq (\ref{radius3}), the
radius can be written as
$R=\alpha_{R}R(0)$,
where
\begin{equation} \label{alphar}
R(0)=\lp\kappa(5-n)f_{0}^{(n)}f_{2}^{(n)}\rp^{\frac{1}{2}}f_{0}^{(n)}\rho_{\rm
c}^{\frac{1-n}{2n}},\hspace{0.3cm} \alpha_{R}\equiv
\lp\frac{f_{0}}{f_{0}^{(n)}}\rp\left[
\frac{f_{0}f_{2}}{f_{0}^{(n)}f_{2}^{(n)}\lp1-\frac{2}{3}\pi \zeta_{\rm
c}f_{0}^{3}(5f_{1}-1)\rp} \right]^{\frac{1}{2}}.
\end{equation}

In table \ref{tablelane2a} we show the values of the enhancement factors $\alpha_{M}$
and $\alpha_{R}$ for different values of $\zeta_{\rm c}$ and different
polytropic index $n$. We also show the values of the critical ratio
$\zeta_{\rm crit}$ which separates the configurations with definite
ratio such that a zero $\xi_{1}$ exist provided that $\zeta_{\rm
c}<\zeta_{\rm crit}$.  We will show some examples where the
enhancement factors may be relevant in section 5.


\begin{table}
\begin{center}
\begin{tabular}{cccc}\hline\hline
 $n$ &$\zeta_{\rm c}=0.1$&$\zeta_{\rm c}=0.05$ &$\zeta_{\rm c}=0.001$\\ \hline 
$1$ &$(1.12,1.29)$ &$(1.05,1.12)$ &$(1.001,1.002)$ \\ 
$1.5$&$(-,-)$&$(1.11,1.17$ &$(1.002,1.003)$ \\ 
$3 $&$(-,-)$ &$(-,-)$ &$(1.022,1.01)$ \\ \hline \hline
\end{tabular}
\caption[Numerical values for the enhancement factors in the
polytropic model]{{Numerical values for the enhancement
factors $(\alpha_{R},\alpha_{M}$) (values for $n=1$ have been taken from
\cite{bala3}). The symbol $-$ indicates a non defined
radius.}}\label{tablelane2a}
\end{center}
\end{table}

\section{The isothermal sphere}
The isothermal sphere is a popular model in astrophysics, either to
model large astrophysical and cosmological objects (galaxies, galaxy clusters)
\cite{lynden,penston,yabu,sommer,chavanis1,more}, to examine the so-called
gravothermal catastrophe
\cite{binney,natarajan,lombardi} and finally to
compare observations with model predictions \cite{rines,more}. 
In the limit $n\to \infty$ in the polytropic equation of state one obtains the description for an
isothermal sphere, (ideal gas configuration) with
\begin{equation} \label{xxx11}
p=\sigma^{2}\rho
\end{equation}
where $\sigma$ is the velocity dispersion ($\sigma^{2}\propto T$).  The pattern we found
in section 3 for $n \ge 5$ gets confirmed here: no finite radius of
the configurations is found with $\Lambda$, the asymptotic behavior is
not $\rho \to 0$ as $r \to \infty$ (but rather $\rho \to 2 \rho_{\rm
vac}$) and, as we will show below, other attempts to define a proper
finite radius are not satisfactory.

The results for a finite value of the index $n$ are
defined in the limit $n\to \infty$ only asymptotically in the case
$\Lambda=0$. We consider this as an acceptable behavior of the
density.  Because of the limiting case $n\to \infty$ the analysis for
the isothermal sphere must be done in a slightly different way.  As
was done in Eq. (\ref{new1}), we can integrate the equilibrium
equations (Euler and Poisson's equations) and obtain an explicit
dependence of the density with $\rvac$ as
\begin{equation}
\label{new3}
\rho(r)=\rho_{c}\exp\left[-\frac{1}{\sigma^{2}}(\Phi_{\rm
grav}(r)-\Phi_{\rm grav}(0))\right]\exp\left[\frac{8}{3\sigma^{2}}\pi
\rvac r^{2}\right],
\end{equation}
The resulting differential equation for the density with cosmological
constant can be written as
\begin{equation}
\label{isothermaljaja}
\frac{\sigma^{2}}{r^{2}}\frac{\dd }{\dd r}\lp r^{2}\frac{\dd \ln \rho
}{\dd r}\rp=-4\pi \rho+8\pi \rvac , \nonumber
\end{equation}

\begin{figure}
\begin{center}
\includegraphics[angle=0,width=9cm]{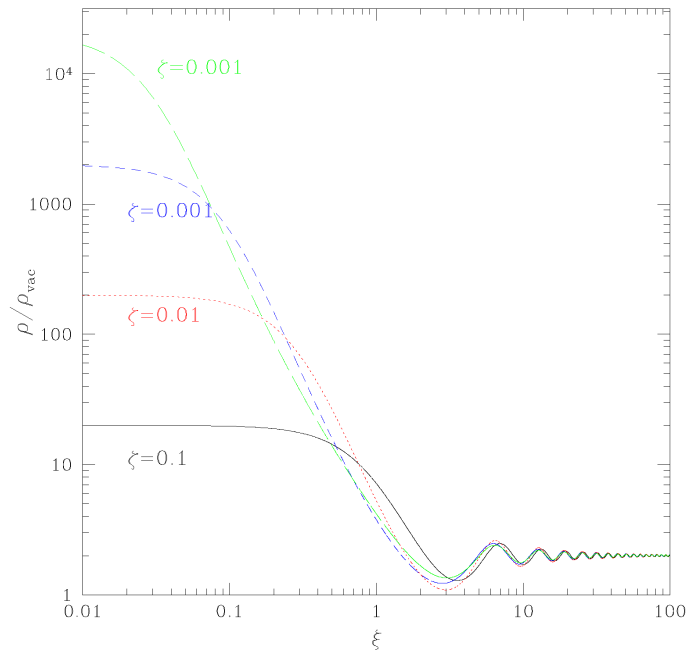}
\caption[Isothermal sphere]{{Scaled density
$\rho/\rvac=e^{\psi}$ for the isothermal sphere at different values of
central density. The solutions oscillate around the value
$\rho=2\rvac$.}}
\label{iso2}
\end{center}
\end{figure}

This differential equation could be treated in the same way as we did
for the polytropic equation of state, i.e, by defining a new function
$\psi\sim \rho/\rho_{\rm c}$, but here we can already use the fact
that the cosmological constant introduces scales of density, length
and time \cite{bala1}. Let us then define the function
$\psi(\xi)=\ln(\rho(r_{0}\xi)/\rvac)$ and $r=r_{0}\xi$, with $r_{0}$
the associated length scale. Since we are now scaling the density with
$\rvac$, the associated length scale $r_{0}$ should be also scaled by
the length scale imposed by $\Lambda$:
\begin{equation}\label{aji}
r_{0}=\sigma\rl= 13.34 \lp\frac{\sigma}{10^{3}\, {\rm km}/{\rm s}}\rp
\,{\rm Mpc}.
\end{equation}
For an hydrogen cloud with $\sigma \sim 4$ km$/$s we have
$r_{0}\approx 40$ kpc which is approximately the radius of an
elliptical (E0) galaxy. 
In terms of the function $\psi(\xi)$, the differential
equation governing the density profile is then written as
\footnote{Compare with Eq. 374 of \cite{chan} or Eq.1 of
\cite{natarajan} where the density is scaled by the central
density. The factor $1$ on the r.h.s of (\ref{isothermal}) is due to
$\rvac$.}
\begin{equation}
\label{isothermal}
\frac{1}{\xi^{2}}\frac{\dd }{\dd \xi}\lp\xi^{2}\frac{\dd \psi }{\dd
\xi}\rp=1-\frac{1}{2}e^{\psi},
\end{equation}
so that according to Eq. (\ref{new3}) we may write
\begin{equation}
\label{new4}
\psi(r/r_{0})=\ln\lp\frac{\rho_{\rm c}}{\rvac}\rp-\sigma^{-2}\lp
\Phi_{\rm grav}(r)-\frac{8}{3} \pi \rvac r^{2}\rp.
\end{equation} 
From this we can derive different solutions $\psi$ depending on the
initial condition $\psi(\xi=0)=\ln (\rho_{\rm c}/\rvac)$ and $\dd \psi(\xi)/\dd \xi=0$ at $\xi=0$. In
Fig. \ref{iso2} we show numerical results for the solutions of
equation (\ref{isothermal}) using different values values of
$\rho_{\rm c}/\rvac$.  As it is the case for $n>5$, the radius cannot
be defined by searching the first zero of the density i.e. the value
$\xi_{1}$ such that $e^{\psi(\xi_{1})}=0$ (including $\psi \to
-\infty$).  In this case, the behavior of the derivative of the
density profile changes as compared with the the $\Lambda=0$ case
since with increasing $\xi$ the density starts oscillating around the
value $\rho=2\rvac$ such that for $\xi\to \infty $ one has a solution
$\rho\to 2\rvac$. This can be checked from (\ref{isothermal}) which
corresponds to the first non trivial solution for $\rho$. 
This behavior implies that there exist a value of $\xi=\xi_{1}$ where
the derivative changes sign and hence the validity of the physical
condition required for any realistic model i.e. $\dd \rho/\dd r <0$
should be given up unless we define the size of the configuration as
the radius at the value of the first local minimum. We will come back
to this option below to show that it is not acceptable.  A second
option would be to set the radius at the position where the density
acquires for the first time its asymptotic value $2\rho_{\rm vac}$.  We
could motivate such a definition by demanding that the density at the
boundary goes smoothly to the background density. This is for tow
reasons, however, not justified.  First, we recall that a positive
cosmological constant leads to a repulsive 'force' as it accelerates
the expansion. A negative cosmological constant could be modeled in a
Newtonian sense by a constant positive density which, however, is
strictly speaking still not a background density.  Secondly, if we
include the background density $\rho_b$ we would have started with
$\rho+\rho_b$ (with a dynamical equation for $\rho$ being
$\rho_b=$constant) in which case the boundary condition would again be
$\rho (R)=0$ to define the extension of the body (or at least, $\rho
\to \infty$ as $r\to \infty$).  Hence, this second option can be
excluded on general grounds.  In any case as can be seen from Figure
\ref{iso2} both definitions would yield two different values of
radius.
Since the first candidate to define a radius is based upon a physical
condition of the configuration, we could expect this definition as the
more suitable one.  However such a definition must be in agreement
with the observed values for masses and radius of specific
configurations and the validity of this definition can be put to test
by the total mass of the configuration, given as
\begin{eqnarray}\label{massiso}
M=
1.55\times 10^{15}\lp\frac{\sigma}{10^{3}\, {\rm km}/{\rm
s}}\rp^{3}f(\xi_{1})\,M_{\odot},\hspace{0.8cm} f(\xi_{1}) &\equiv
&\int_{0}^{\xi_{1}}\xi^{2}e^{\psi}\dd\xi.
\end{eqnarray}
Combining (\ref{aji}) and (\ref{massiso}) we can write
\begin{equation}\label{massiso2}
M=653.55\lp\frac{R}{{\rm
kpc}}\rp^{3}\xi_{1}^{-3}f(\xi_{1})\,M_{\odot}.
\end{equation}
If we define the radius at the first minimum (see Fig. \ref{iso2}), we
find $M\approx 2\times 10^{9}\lp R/{\rm
kpc}\rp^{3}M_{\odot}$. Although this might set the right order of
magnitude for the mass of a E0 galaxy if we insist on realistic values
for the respective radius, say $R\sim 10$ kpc, the picture changes
again as the radius is fixed by (\ref{aji}) which gives $R\approx
5.3\times 10^{6}\lp\sigma/10^{3}{\rm km/s}\rp {\rm kpc}$.  In order to
get a radius of the order of kpc with masses of the order of
$10^{10}M_{\odot}$ we would require $\sigma \sim 10^{-5}{\rm
km/s}$. This differs by almost eight orders of magnitude with the
measured values for the velocity dispersion $\sigma$ in elliptical
galaxies ($\sigma \sim 300$ km$/$s) or with the Faber-Jackson Law for
velocity dispersion \cite{padma}. We conclude that defining the radius by the position of the first minimum is not a realistic solution. As the last option to get realistic values for the parameters of the
configuration we consider the brute force method to simply fix the
value of $\xi_{1}$.  It is understood, however, that this method is
not acceptable if we insist that the model under consideration has
some appealing features (without such features almost any model would
be phenomenologically viable).  Therefore we discuss this option only
for completeness.  For a configuration with $\rho_{\rm
c}=10^{5}\rho_{\rm vac}$, say an elliptical galaxy, we fix the radius
at $R\sim 50$ kpc in Eq.(\ref{aji}) and using a typical value
for the velocity dispersion $\sigma \sim 300$ km$/$s we get
$\xi_{1}\sim 0.012$ which implies $f(\xi_{1})\sim 0.036$. The mass
in Eq.(\ref{massiso}) is then given as $M\sim 1.5\times 10^{12}M_{\odot}$
while the density at the boundary is $\rho_{R}\sim 36000 \rho_{\rm
vac}$, that is, $\rho_{\rm c}\sim 2.35 \rho_{R}$. In table
\ref{tablelane2xx} we perform the same exercise for other radii.  The
resulting mean density is in accordance with the observed values of
the mean density of astrophysical objects ranging from an small
elliptical galaxy to a galaxy cluster.  However, as mentioned above,
the model introduces an arbitrary cut-off and cannot be considered as
a consistent model of hydrostatic equilibrium.

In summary, the attempts to define a finite radius for the isothermal
sphere fail in the presence of a cosmological constant either because
such a model fails to reproduce certain phenomenological values (if
the definition of the radius is fixed by the first minimum) or because
the definition is technically speaking quite artificial to the extent
of introducing arbitrary cut-offs. Note that this conclusion is valid 
almost for any object as the density of the isothermal sphere
with $\Lambda$ has a minimum whatever the central density we choose.

\begin{table}
\begin{center}
\begin{tabular}{ccccc}\hline\hline 
 $R/{\rm kpc}$ &$\xi_{1}$ & $\rho_{\rm c}/\rho(R)$ & $M/M_{\odot}$&
$\bar{\rho}$ (gr/cm$^{3}$) \\ \hline $10$ &$0.00249$ &$1.0005$
&$2.17\times 10^{8}$ & $3.4\times 10^{-25}$ \\ $50$ &$0.01248$
&$1.0129$ &$2.7\times 10^{10}$ & $1.94\times 10^{-25}$ \\ $100$
&$0.0249$ &$1.052 $ &$2.11\times 10^{11}$ & $7.2\times 10^{-26}$ \\
\hline \hline
\end{tabular}
\caption[Numerical values ]{{Values for $\xi_{1}$,
$\rho_{\rm c}/\rho(R)$ and the mass for different values of radius and
for $\rho_{\rm c}=10^{3}\rvac$  with $\sigma=300$km/s.}  }\label{tablelane2xx}
\end{center}
\end{table}

\section{Exotic astrophysical configurations}
In this section we will probe into the possibilities of exotic, low
density configurations.  The global interest in such structures is
twofold. First, the cosmological constant will affect the properties
of low density objects. Secondly, $\Lambda$ plays effectively the role
of an external, repulsive force.  Hence a relevant issue that arises
in this context is to see whether the vacuum energy density can partly
\emph{replace} the pressure which essentially is encoded in the
parameter $\kappa$ ($p=\kappa \rho^{\gamma}$).  By the word
\emph{replace} we mean that we want to explore the possibility of a
finite radius as long as the pressure effects are small in the
presence of $\rvac$.

\subsection{Minimal density configurations}
\begin{figure}
\begin{center}
\includegraphics[angle=0,width=9cm]{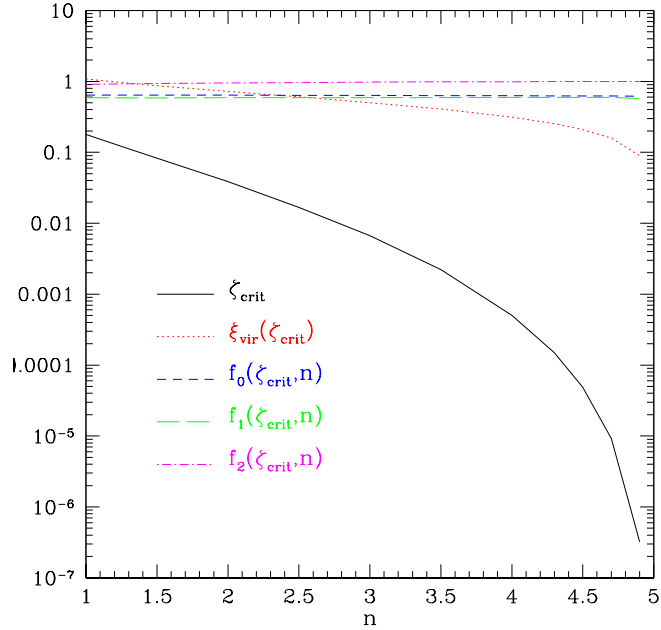}
\caption[]{{Same as fig. \ref{tablelane2}, now the functions $f_{i}$ being evaluated at the value $\xi_{\rm vir}$ solution of Eq.(\ref{virtras}).}}
\label{lanecrit}
\end{center}
\end{figure}

As mentioned above the effect of a positive cosmological constant  on
matter is best understood as an external repulsive force. In previous
sections we have probed into one extreme which describes the situation
where a relatively low density object is pulled apart by this force (to an extent that
we concluded that the isothermal sphere is not a viable model in the presence of
$\Lambda$). Limiting conditions when this happens were derived. 
On the other hand, approaching with our parameters these limiting conditions, but remaining
still on the side of equilibrium, means that relatively low density objects can be still
in equilibrium thanks to the positive cosmological constant.
The best way to investigate low density structures is to to use the
lowest possible central density. As explained in section 3, for every
$n$ there exist a ${\cal A}_n$ such that $\rho_c \ge {\cal A}_n
\rho_{\rm vac}$ which defines the lowest central density. Certainly, a
question of interest is to see what such objects would look like. 
We start with the parameters of the configuration. The radius at the critical value $\zeta_{\rm crit}$ is given by Eq.(\ref{radius3}) 
after taking the limit  $\rho_{\rm c}\to \maa_{n}\rvac$ (see (\ref{An})). It is given by 
\begin{equation}
\label{rque}
R_{\rm crit}=2.175\lp\frac{M}{10^{12}M_{\odot}}\rp^{1/3}f_{0}(\zeta_{\rm crit};n) \zeta_{\rm crit}^{1/3}\,\,\rm Mpc.
\end{equation}
From fig. \ref{lanecrit} we can see that the product $f^{3}_{0}(\zeta_{\rm crit};n)\zeta_{\rm crit}$ changes a little round the value $\sim 0.2$ as we change the index $n$, so that these polytropic configurations will have roughly the same radius (for a given mass). 
This implies that such configurations have approximately the same average density $\bar{\rho}(\zeta_{\rm crit})=1.64 \rho_{\rm crit} \approx 2.34 \rho_{\rm vac}$. That is, such configurations have a mean density of the order of the of the critical density of the universe.
Given such a density we would, at the first glance, suspect that the object described by this
density cannot be in equilibrium. However, our result follows strictly from hydrostatic
equilibrium and therefore there is no doubt that such object can theoretically exist. Furthermore,
$\bar{\rho}$ satisfies the inequalities derived in \cite{nowakowski1} and \cite{bala2} from virial
equations and Buchdahl inequalities which guarantee that the object is in equilibrium ($\bar{\rho} > 2\rho_{\rm vac}$).
Interestingly, the central density for such objects has to be much higher than $\bar{\rho}$ as, e.g., for $n=3$ we have
$A_3 \approx 300$ and therefore $\rho_c > 300 \rho_{\rm vac}$.
Note that these values have been given from the solution of the Lane-Emden equation, 
which is a consequence of dynamical equations reduced to describe our system in a steady state. 
However, we have not tried to solve explicitly quantities from the virial theorem. 
This makes sense as for Dark Matter Halos (DMH) the parametrized density profiles go often only
asymptotically to zero and the radius of DMH is defined as a virial radius where the density
is approximately two hundred times over the critical one. Therefore, an analysis 
using virial equations seems to be adequate here.
For constant density Eq.(\ref{vpol}) can be expressed as a cubic equation \cite{bala1} 
for the radius at which the virial theorem is satisfied (let us ignore for these analysis 
any surface terms coming from the tensor virial equation). However, 
if the density is not constant, this expression becomes a transcendental equation for 
the dimensionless radius $\xi_{\rm vir}=R_{\rm vir}/a$. This equation is
\begin{equation}\label{virtras}
\xi_{\rm vir}^{2}=\frac{6(5-n)f_{0}^{3}}{(n+1)\left[3-2\pi \zeta_{\rm c}(5f_{1}-1)f^{3}_{0}\right]},
\end{equation}
understanding the functions $f_{i}$ now as integrals up to the value $\xi_{\rm vir}$. Once we fix $\zeta_{\rm crit}$ for a given index $n$, we use as a first guess for the iteration process the value $\xi_{1}(\zeta_{\rm crit})$. In fig \ref{lanecrit} we show the behavior of the functions $f_{i}$ and the solutions of Eq.(\ref{virtras}). For these values, Eq. (\ref{rque}) gives for $n=3$ a radius 
\begin{equation}\label{xxx14}
R_{\rm vir}=25.8 \lp\frac{M}{M_{\odot}}\rp^{1/3}\,\rm pc,
\end{equation}
which can be compared with the radius-mass relation derived in the top-hat sphericall collapse \cite{padma}
\begin{equation}\label{contrast1}
R_{\rm vir}=21.5h^{-2/3}(1+z_{\rm vir})^{-1}\lp\frac{M}{M_{\odot}}\rp^{1/3}\,\rm pc,
\end{equation}
where $h$ is the dimensionless Hubble parameter and $z_{\rm vir}$ is the redshift of virialization. The resulting average density is then of the order of the value predicted by the top-hat sphericall model:
\begin{equation}\label{contrast}
\bar{\rho}=\lp\frac{3\Omega_{\rm vac}}{2\pi \zeta_{\rm crit}f_{0}^{3}}\rp\rho_{\rm crit}\approx 200 \rho_{\rm crit}, 
\end{equation}
Note with the help of Eq.(\ref{a}) and (\ref{rque}) that the mass can be written as proportional to the parameter $\kappa^{3/2}$ (introduced in the polytropic equation of state Eq.(\ref{poly})).  Therefore $\kappa \to 0$ is equivalent to choosing a small pressure and, at the
same time, a small mass which, in case of a relatively small radius, amounts to
a diluted configuration with small density and pressure.
Without $\Lambda$ such configurations would be hardly in equilibrium.
Hence, for the configuration which has the extension of pc, Eq.(\ref{xxx14}), with one solar mass we conclude that the equilibrium is not fully due to the pressure, but partially maintained also by $\Lambda$. This is possible, as $\Lambda$ exerts an
outwardly directed non local force on the body. Other mean densities, also independent of $M$ are for $n=1.5$ and $n=4$, respectively:
\begin{equation} \label{xxx34}
\bar{\rho}= 15.3\rho_{\rm crit},\hspace{1cm}\bar{\rho}= 2.6 \times 10^{3}\rho_{\rm crit},
\end{equation}
The first value is close to $\rho_{\rm crit}$ and therefore also to $\rho_{\rm vac}$. Certainly,
if in this example we choose a small mass, equivalent to choosing a
negligible pressure, part of the equilibrium is maintained by the
repulsive force of $\Lambda$.  In \cite{bala2} we found a simple
solution of the hydrostatic equation which has a constant density of
the order of $\rho_{\rm vac}$. The above is a non-constant and
non-trivial generalization of this solution.

\subsection{Cold white dwarfs}
The neutrino stars which we will discuss in the subsequent subsection
are modeled in close analogy to white dwarfs. Therefore it makes sense
to recall some part of the physics of white dwarfs. In addition we can
contrast the example of white dwarfs to the low density cases affected
by $\Lambda$.
 
In the limit where the thermal energy $k_{\rm B}T$ of a (Newtonian)
white dwarf is much smaller than the energy at rest of the electrons
($p_F $, these configurations can be treated as polytropic
configuration with $n=3$.  This is the ultra-relativistic limit where
the mass of electrons is much smaller than Fermi's momentum $p_{\rm
F}$. In the opposite case we obtain a polytrope or configuration with
$n=3/2$.  \cite{weinberg,shapiro}. In both cases, the parameter
$\kappa_{n}$ from the polytropic equation of state is given as
\begin{equation} 
\label{kn}
\kappa_{3}=\frac{1}{12\pi
^{2}}\lp\frac{3\pi^{2}}{m_{n}\mu}\rp^{\frac{4}{3}}
,\hspace{1cm}\kappa_{3/2}=\frac{1}{15
m_{e}\pi^{2}}\lp\frac{3\pi^{2}}{m_{n}\mu}\rp^{\frac{5}{3}},
\end{equation}
where $m_{n}$ is the nucleon mass, $m_{e}$ is the electron mass and
$\mu$ is the number of nucleons per electron.  Using the Newtonian
limit with cosmological constant, we can derive the mass and radius of
these configurations in equilibrium.  In the first case, for $n=3$ the
mass is written using (\ref{fmas2aa}) as $M_0=\mathcal{G}(n=3)$ which
corresponds approximately to the Chandrasekhar's limit (strictly
speaking a configuration would have the critical mass, i.e, the
Chandrasekhar's limit, if its polytropic index $\gamma$ is such that
$\gamma=\gamma_{\rm crit}$).  For this situation one has $M_{0}(n=3)=
5.87 \mu^{-2}M_{\odot}$ and $R_{0}(n=3)= 6.8
(\bar{\rho}_{\odot}/\rho_{\rm c})^{1/3}\mu^{-\frac{2}{3}} R_{\odot}$.
On the other hand, for $n=3/2$ one obtains $M_{0}(n=3/2)=3.3\times
10^{-3}(\bar{\rho}_{\odot}/\rho_{\rm c})^{-1/2}\mu^{-5/2}M_{\odot}$
and the radius is given by
$R_{0}(n=3/2)=0.27(\bar{\rho}_{\odot}/\rho_{\rm
c})^{1/6}\mu^{-\frac{5}{6}}R_{\odot}$ where $\bar{\rho}_{\odot}$ is
the mean density of the sun.  Since for these configurations the ratio
$\zeta_{\rm c}$ is much smaller than $10^{-4}$ we see from Fig.
\ref{tablelane2} that the effects of $\Lambda$ are almost negligible.
The critical value of the ratio $\zeta_{\rm c}$ gives for $n=3$ the
inequality $\rho_{\rm c}>307.69\rho_{\rm vac}$ and for $n=3/2$ the
same limit reads $\rho_{\rm c}>24.24\rho_{\rm vac}$.  Central
densities of white dwarfs are of the order of $10^{5}{\rm gr}/{\rm
cm}^{3}$ which corresponds to a deviation of nearly thirty orders of
magnitude of $\rho_{\rm vac}$.

\subsection{Neutrino stars}
An interesting possibility is to determine the effects of $\rho_{\rm
vac}$ on configurations formed by light fermions. Such configurations
can be used, for instance, to model galactic halos
\cite{dolgov,lattanzi,jetzer, boerner}. While discussing the phenomenological
interest of fermion stars below, we intend to describe such a halo.
Clearly, these kind of systems will maintain equilibrium by
counterbalancing gravity with the degeneracy pressure as in a white
dwarf.  For stable configurations, i.e, $n=3/2$, one must replace the
mass of the electron and nucleon by the mass of the considered fermion
and set $\mu=1$ in (\ref{kn}) and (\ref{fmas2aa}).  We then get for
the mass and the radius:
\begin{eqnarray}
\label{mrar}
M_{0}&=&3.28\times
10^{28}\lp\frac{\rho_{c}}{\bar{\rho}_{\odot}}\rp^{\frac{1}{2}}\lp\frac{{\rm
eV}}{m_{f}}\rp^{4}\, M_{\odot} \,\,=\,\,3.64\times 10^{14} \zeta_{\rm
c}^{-1/2}\lp\frac{{\rm eV}}{m_{f}}\rp^{4}\, M_{\odot}, \\ \nonumber
R_{0}&=&1.31\times
10^{-4}\lp\frac{\rho_{c}}{\bar{\rho}_{\odot}}\rp^{-\frac{1}{6}}\lp\frac{{\rm
eV}}{m_{f}}\rp^{\frac{4}{3}}\, {\rm Mpc}=6.16 \zeta_{\rm
c}^{1/6}\lp\frac{{\rm eV}}{m_{f}}\rp^{\frac{4}{3}}\, {\rm Mpc}.
\end{eqnarray}
On the other hand, for $n=3$ one has
\begin{eqnarray}
\label{mrara}
M_{0}&=&5.16\times 10^{18}\lp\frac{{\rm eV}}{m_{f}}\rp^{2}\,
M_{\odot}, \\ \nonumber R_{0}&=&0.14\lp\frac{{\rm
eV}}{m_{f}}\rp^{\frac{2}{3}}\lp\frac{\rho_{\rm
c}}{\bar{\rho}_{\odot}}\rp^{-\frac{1}{3}}\, {\rm pc}\,\,=\,\,
3.1\times 10^{5} \zeta_{\rm c}^{1/3}\lp\frac{\rm eV}{m_{\rm
f}}\rp^{2/3} \,{\rm pc},
\end{eqnarray}
The cases represent, among other, possible cosmological configurations
when the fermion mass if of the order of eV, for instance, massive
neutrinos. 
\begin{figure}
\begin{center}
\includegraphics[angle=0,width=10cm]{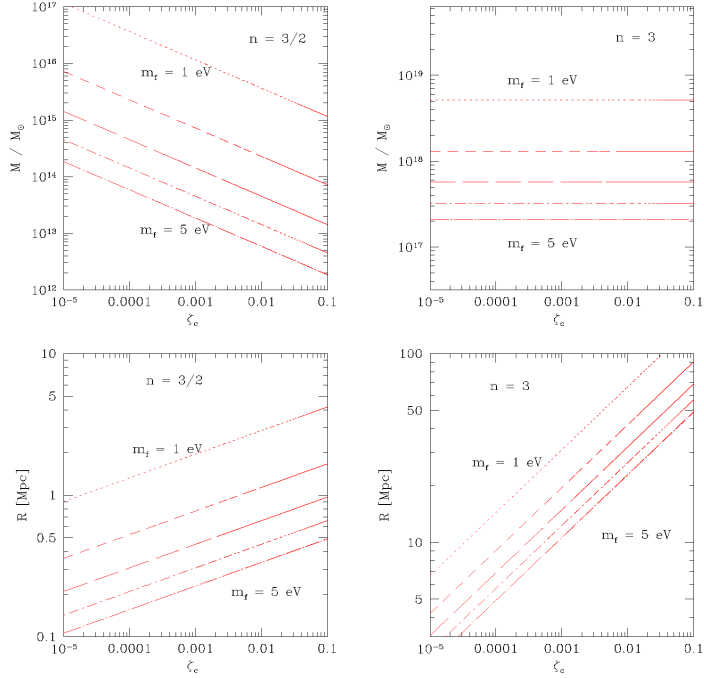}
\caption[]{Mass and radius for different central
densities, for index $n=3$ and $n=3/2$. The masses range from $m_{f}=1$ eV up to $m_{f}=5$ eV.}\label{tablelane244}
\end{center}
\end{figure}
In figure \ref{tablelane244}
some representative values for the choice $m_f=1$ eV up to   $m_f=5$ eV are
given. Obviously in the case $n=1.5$ the values for the mass and
radius are sensitive to the choice of $m_f$.  Indeed, the dependence
on the fermion mass is much stronger than on the central density.  It is justified to speculate
that a relative low density objects, affected by $\Lambda$, might
exist.  If then, as in an example we choose $m_f \sim 5$ eV and
$\rho_c \sim 40 \rho_{\rm vac}$ then the mass comes out as $10^{12}$
solar masses with a radius of the order of magnitude of half Mpc which
might indeed be the dark matter halo of a galaxy (or at least part of
the halo).  As pointed out before, such configurations must have a
central density greater than $24.4 \rho_{\rm vac}$ in order to be in
equilibrium. From table \ref{tablelane2a} we see that the effect on a
configuration with $\zeta_{\rm c}\sim 0.05$ is represented in an
increase in the mass by $17\%$ with respect to $M_{0}$ and an
increment of $11\%$ in the radius. Then the conclusion would be that
$\Lambda$ affects such a dark matter halo.  This is to be taken with
some caution as the fermions in such a configuration would be
essentially non-relativistic.  Note also that it is not clear if
neutrinos make up a large fraction of the halo, however, we can also
speculate about a low density clustering around luminous matter. Of
course, allowing larger central density might change the picture.
However, the emerging scenario would not necessarily be a viable
phenomenological model. For instance, changing the value of $m_f$ from
eV to keV (MeV) would reduce the mass by twelve (twenty four) orders
of magnitude which is definitely too small to be of interest. We could
counterbalance this by increasing the central density by twenty four (
forty eight!)  orders of magnitude. Such a 'countermeasure' would,
however, result in a reduction of the radius by eight (sixteen!)
orders of magnitude, again a too small length scale to be of
importance for dark matter halos. In other words, the example with a
fermion mass of the order of one eV and low central density is
certainly of some phenomenological interest.

The case $n=3$ is similarly stringent. A neutrino mass of $1-5$ eV
gives a mass for the entire object of the order of ten to the eighteen
solar masses which is too large. A fermion mass of several keV would
be suitable for a galaxy halo (a mass of the order Me V and higher
would give a too small total mass). With a relative low central
density as before (see Figure \ref{tablelane244}) we then obtain the
right order of magnitude for the halo. But then we will have to live
with the fact that such halo reaches up to the next large galaxy.  Briefly, we touch upon the other possible application of fermions stars
which have been discussed as candidates for the central object in our
galaxy. If we allow the extension of this object to be $120$ AU and
the mass roughly $2.6$ million solar masses, then the fermion mass
would come out as $10^{4}$ eV for $n=1.5$ ($10^6$ eV for $n=3$) and
the central density as $10^{22}\rho_{\rm vac}$ ($10^{28}\rho_{\rm
vac}$ for $n=3$).

\subsection{Boson stars}
We end the section by putting forward a speculative question in
connection with boson stars.  The latter are general relativistic
geons and can be treated exactly only in general relativistic
framework i.e.  these kind of configurations are based on the
interaction of a massive scalar field and gravitation which leads to
gravitational bounded systems.  These objects have been also widely
discussed as candidates for dark matter \cite{ruffini,chi}.  On the
other hand, variational methods in the connection with the
Thomas-Fermi equations give relatively good results even without
invoking the whole general relativistic formalism.  By including
$\Lambda$ we essentially introduce into the theory a new scale, say in
this case a length scale $r_{\Lambda}=1/\sqrt{\Lambda}$.  The basic
parameters of dimension length in a theory with a boson mass $m_B$ and
a total mass $M=N_Bm_B$ where $N_B$ is the number of bosons are (for a
better distinction of the  different length scales, we restore in
this subsection the value of $G_N$)
\begin{equation} \label{L}
L_1=r_s=G_NM, \,\, L_2=r_B=\frac{1}{m_B}, \,\,
L_3=r_{\Lambda}=\frac{1}{\sqrt{\Lambda}}.
\end{equation}
The resulting radius of the object's extension can be a combination of
these scales i.e.
\begin{eqnarray} \label{R}
R_i^{(1)}&\propto &L_i, \,\, R_i^{(2)}\propto N_B^nL_i \nonumber \\
R_1&\propto &(L_i^2L_j)^{1/3},\,\, R_2\propto N_B^nR_1
\end{eqnarray}
and similar combination of higher order. Which one of the combination
gets chosen, depends on the details of the model.  In a close analogy
to \cite{spruch,eckehard} we can examine this taking into account the
presence of a positive cosmological constant by considering the energy
of such configuration as a two variable function of the mass and the
radius $E=E(R,M)$
\begin{equation}
E\sim \frac{N_{\rm B}}{R}-\frac{G_{\rm N}m_{\rm B}^{2}N_{\rm
B}^{2}}{R}+\frac{8}{3}\pi G_{\rm N}\rvac m_{\rm B}N_{\rm B}R^{2}
\end{equation}
The first term corresponds to the total kinetic energy written as
$\mak=N_{\rm B}p=N_{\rm B} /\lambda$ and taking $\lambda\sim R$.  The
second term is the gravitational potential energy and the third term
corresponds to the contribution of the background (see
Eq. (\ref{virialeq})).  By treating mass and radius as independent
variables (we think this this is the right procedure since the radius
will depend on the 'external force' due to $\rvac$), we extremize the
energy leading to the following values of mass and radius:
\begin{equation}
M\sim \frac{1}{G_{\rm N}m_{\rm B}}\sim 10^{-10}\lp\frac{\rm eV}{m_{\rm
B}}\rp M_{\odot},
\hspace{1cm}R\sim \lp \frac{1}{8\pi G_N m_{\rm B}\rvac}\rp^{1/3}\sim
10^{5}\lp\frac{\rm eV}{m_{\rm B}}\rp^{\frac{1}{3}}\,R_{\odot}
\end{equation}
Such values would lead to a mean density of the order of
$\bar{\rho}\sim 6\rvac$ i.e. an extremely low density
configuration. The mass given in the last expression is the so-called
Kaup limit \cite{spruch}. Of course, this relative simple treatment
does not guarantee that the full, general relativistic treatment, will
give the same results.  Therefore we consider it as a
conjecture. However, it is also obvious from the the discussion above
that low density boson stars are a real possibility worth pursuing
with more rigor (we intend to do so in the near future).

\subsection{A comparison between $\Lambda$LE profiles and Dark Matter Halos profiles}
It is of some importance to see whether our results from the examination of polytropic hydrostatic
equilibrium or from the virial equations can be applied to Dark matter configurations.
N-body simulations based in a $\Lambda$CDM model of the universe show that the density profile of virialized Dark Matter Halos (DMH) can be described by a profile of the form \cite{nfw} 
\begin{equation}
\rho(r)=(2)^{3-m}\rho_{s}(r/r_{s})^{-m}\lp1+r/r_{s}\rp^{m-3},
\end{equation}
where  $r_{s}$ is the characteristic radius (the logarithmic slope is $\dd \ln \rho /\dd \ln r_{s}=-m+\frac{1}{2}(m-3)$), $\rho_{s}=\rho(r_{s})$ and the index $m$ characterizes the slope of the profile in the central regions of the halo. The mass of the configuration enclosed in the virial radius $r_{\rm vir}$ is given as $M_{\rm vir}=4\pi(2)^{3-m}\rho_{s}r_{s}^{3}F(c)$ such that one can write $\rho_{s}=(1/3)(2)^{m-3}\Delta_{\rm vir}\rho_{\rm crit}c^{3}F^{-1}(c)$, where $c=r_{\rm vir}/r_{s}$ is the concentration parameter, $\Delta_{\rm vir}$ is the ratio between the mean density at the time of virialization and the critical density of the universe ($\Delta_{\rm vir}\approx 18\pi^{2}$ in the top-hat model for a flat Einstein-deSitter universe, while $\Delta_{\rm vir}\sim 104$ in the $\Lambda$CDM cosmological model \citep{diemand}) 
and $F(c)=\int_{0}^{c}x^{2-m}(1+x)^{m-3}\dd x$. This \emph{universal profile} has been widely used in modeling DMH in galaxy clusters, and the comparison of these models with a stellar polytropic-like profile -without the explicit contribution from the cosmological constant in the Lane-Emden equation- can be found in \cite{sussman,arieli}. 

Some differences can be described between the $\Lambda$LE and the NFW profiles. First, on fundamental 
grounds, it is clear that the NFW profile does not satisfy the LE equation; 
a basic reason for this is that dark matter is assumed to be collisionless and is only affected by gravity. 
However, the fact that the real nature of dark matter is still an unsolved issue leaves an open door through which one can introduce interaction between dark matter particles leading to a different equation of state (see for instance \citep{rem}). On functional forms, one sees that at the central region the difference is abrupt, since the Lane-Emden equation has a flat density profile at $r=0$, while the NFW profile has a cuspy profile of the form $\rho \sim r^{-m}$. 
It has been widely discussed how such cuspy profile is inconsistent with data 
showing central regions of clusters with homogeneous cores (see discussion at the
end of this section). Such small slope in the inner regions can be reproduced by the \emph{universal profile} for the case $m=0$, and for the profile derived from the $\Lambda$LE equation (which is just a consequence of initial conditions, and hence, its independent of $\Lambda$). In this case, the effects of the cosmological constant can be reduced to explore the outer regions of the halos and compare, for instance, the slope of the profiles and the virial radius predicted by each one. 

\begin{figure}
\begin{center}
\includegraphics[angle=0,width=12cm]{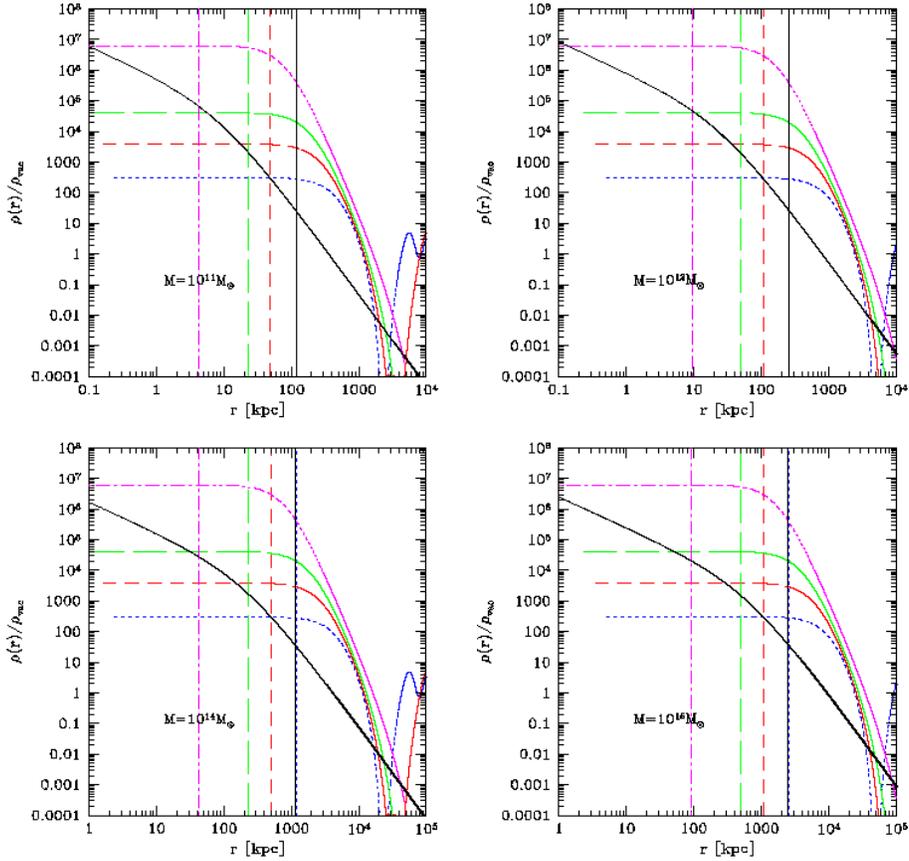}
\caption[]{{NFW profile (black, solid line) compared to the solutions of $\Lambda$LE equation for different masses and different index $n$ ranging from $n=3$(blue,dots), $n=4$ (red,short-dashed line), $n=4.5$ (green,long-dashed line) and $n=4.9$ (magenta,dot short-dashed line), in the limiting case $\zeta_{\rm c}=\zeta_{\rm crit}$. The solution from LE equation is written for the critical value of the parameter $\zeta_{\rm crit}$, shown in table \ref{tablelane2}. 
The vertial black line represents the virial radius from the NFW profile. 
The vertical colored lines represent the virial radius $R_{\rm vir}(\zeta_{\rm crit})=a\xi_{\rm vir}$ for the different polytropic indices.}}\label{compav}
\end{center}
\end{figure}
\begin{figure}
\begin{center}
\includegraphics[angle=0,width=12cm]{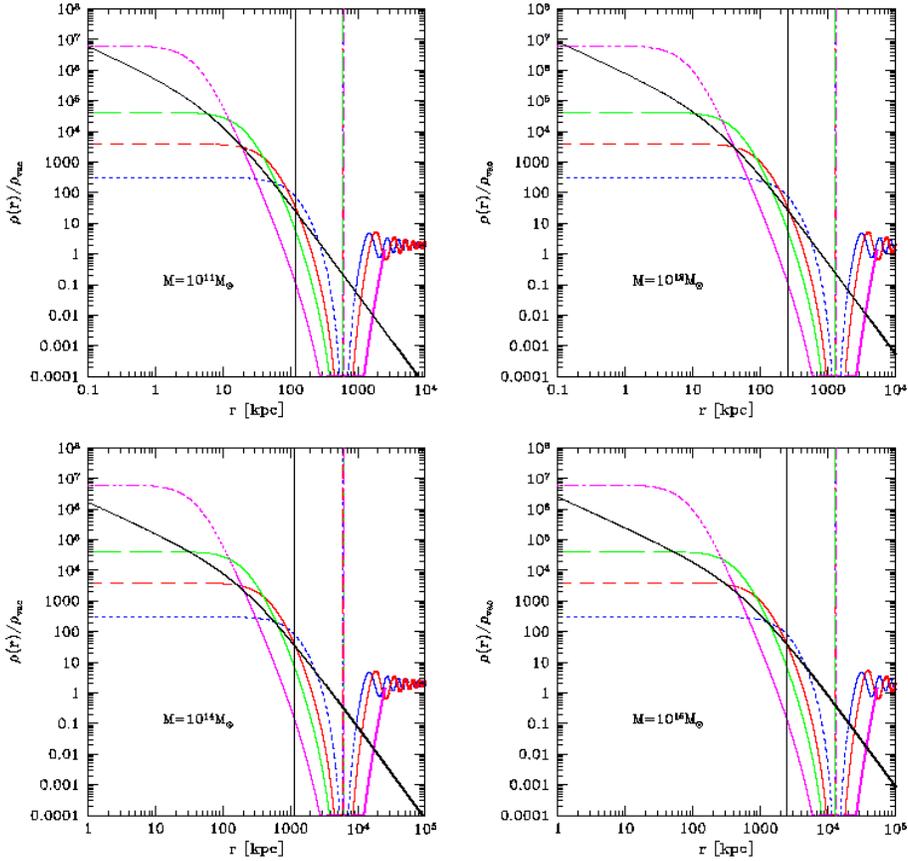}
\caption[]{{Same as fig \ref{compav} but for the radius given by the condition $\psi(\xi_{1})=0$.}} \label{compac}
\end{center}
\end{figure}
\begin{figure}
\begin{center}
\includegraphics[angle=0,width=12cm]{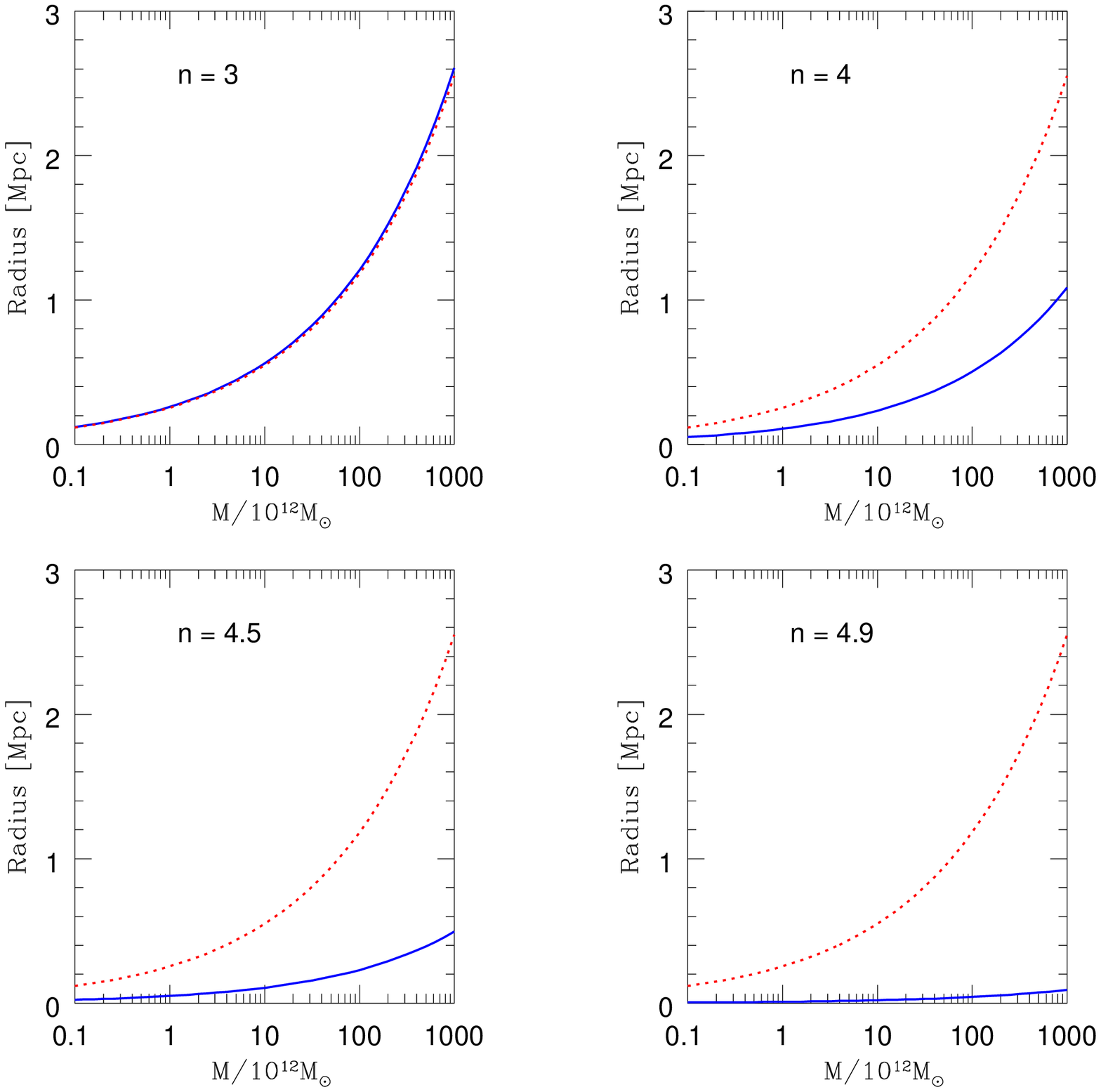}
\caption[]{{Mass-radius relation for the NFW (red, dashed line) profile compared with the solutions Eq. \ref{virtras} (blus, solid line) for four different values of $n$.}} \label{raddd}
\end{center}
\end{figure}

The density profiles predicted by the $\Lambda$LE equation and the NFW profile (for $m=1$) are presented in fig.\ref{compav} for the virial radius given by $\xi_{\rm vir}$ and in fig.\ref{compac} for the radius given by $\xi_{1}$, both with $\zeta_{\rm c}=\zeta_{\rm crit}$. Also, the behavior of the radius for different values of mass and polytropic indices are given in Fig \ref{raddd} (with $r_{s}\approx 25.3\lp M_{\rm vir}/10^{12}M_{\odot}\rp^{0.46}$ kpc and $r_{\rm vir}=1.498\Delta_{\rm vir}^{-1/3}\lp M_{\rm vir}/ 10^{12}M_{\odot}\rp^{1/3}$ Mpc $\approx 255\lp M_{\rm vir}/ 10^{12}M_{\odot}\rp^{1/3}$ kpc \cite{gentile}). 

We see that the virial radius given by the $\Lambda$LE equation is
surprisingly close to the virial radius given by the NFW profile for
$n=3$ in the range of masses shown in fig \ref{raddd}, and as
pointed in Eq.(\ref{contrast}), this value yields for this model
$\Delta^{\Lambda LE}_{\rm vir}\approx 200$. However, as can be seen
from the plots, the $\Lambda$LE density profile is almost flat until
the virial radius. On the other hand, the $m=0$ profile allow us to
parameterize it as $\rho(r)/\rho_{\rm vac}=2\zeta_{\rm
c}^{-1}(1+r/r_{s})^{-3}$, where the concentration parameter $c$ can be
written as $c+1\approx 1.11 \zeta_{\rm c}^{1/3}\Delta_{\rm
vir}^{1/3}$.

\begin{figure}
\begin{center}
\includegraphics[angle=0,width=12cm]{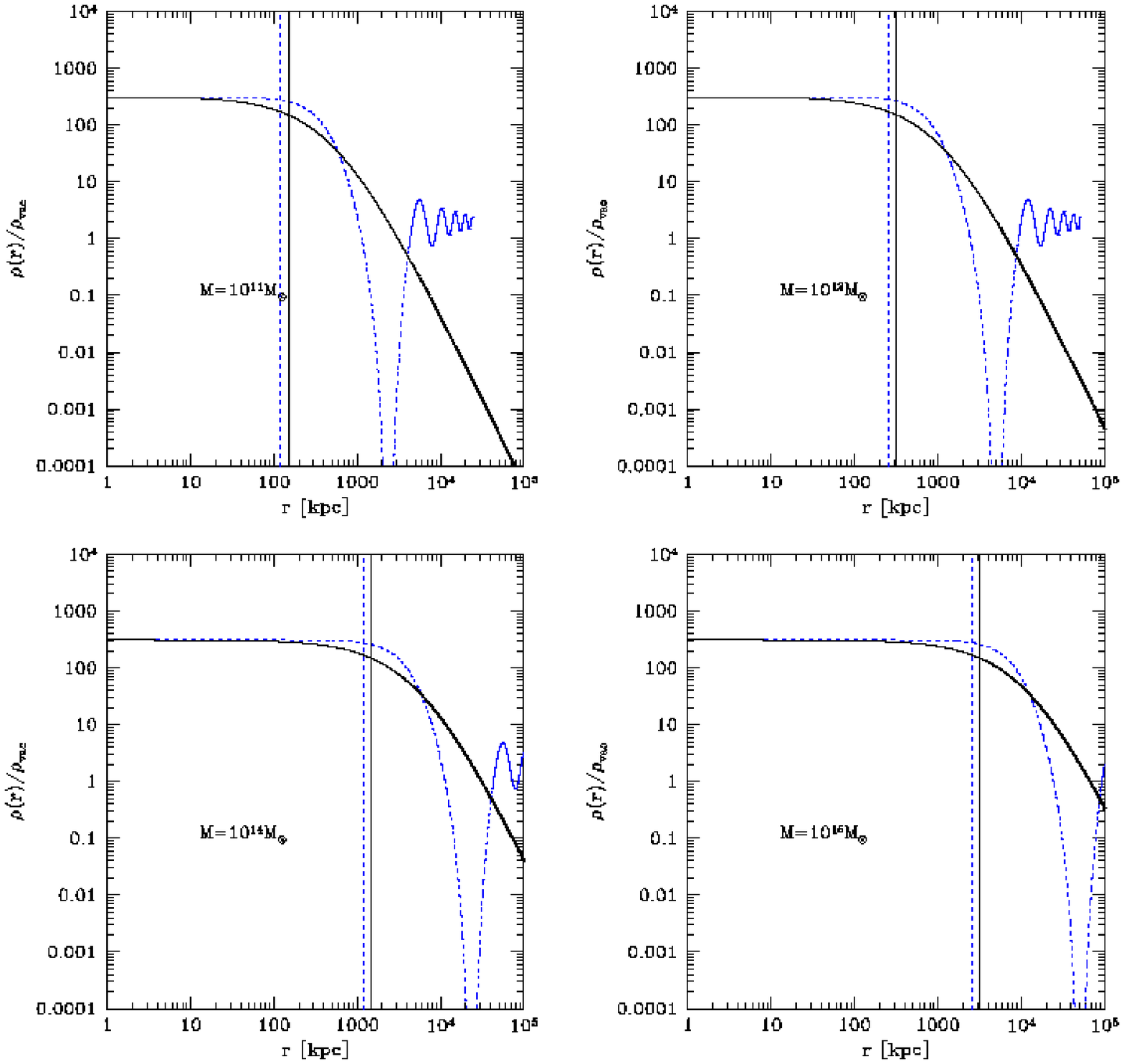}
\caption[]{{Generalized NFW profile (black solid line) for $m=0$ compared with the and $n=3$ $\Lambda$LE profile, with its corresponding $\zeta_{\rm crit}$, for different masses.}} \label{comp1}
\end{center}
\end{figure}
\begin{figure}
\begin{center}
\includegraphics[angle=0,width=12cm]{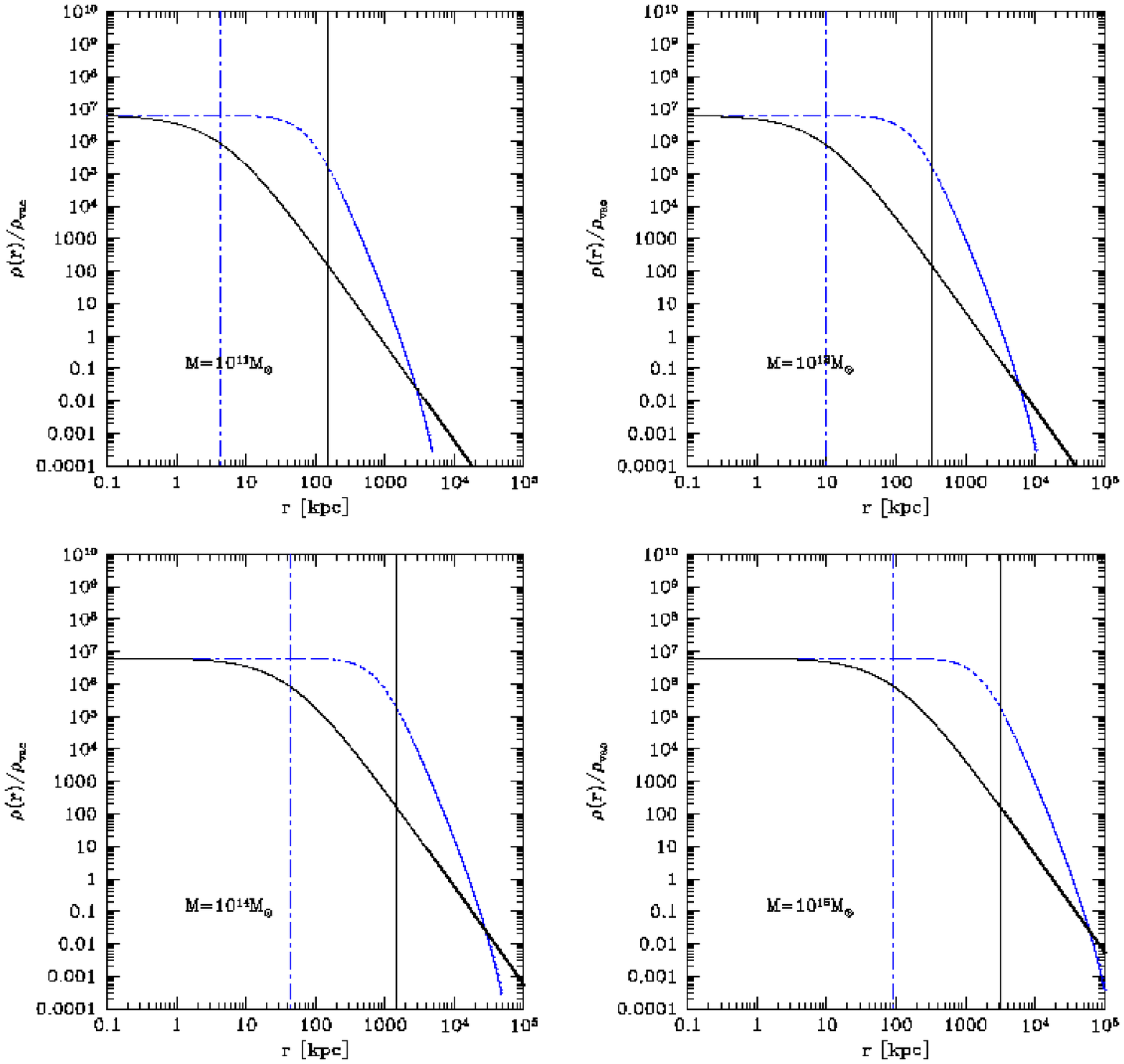}
\caption[]{{Same as fig \ref{comp1} for $n=4.9$}} \label{comp4}
\end{center}
\end{figure}
In fig.\ref{comp1} and \ref{comp4} we have compared both profiles
with the corresponding $\zeta_{\rm crit}$ for  polytropic
index $n=3$ and $n=4.9$.  For the first case we see that the virial radius are of the
same order of magnitude than the one predicted by the NFW profile. For
$n=4.9$, these quantities differ by one order of magnitude. In all
cases, the slope of the NFW profiles changes faster than the
$\Lambda$LE profile, which implies that the polytropic configurations
enclosed by $r_{\rm vir}$ display almost a constant density.

The comparison we made above between the polytropic configuration and the NFW profiles shows that
up to the cuspy behaviour the polytropic results agree with NFW for the polytropic index $n=3$.
It is worth pointing out here that it is exactly this cuspy behaviour which seems to be at odds
with observational facts \cite{arieli, Matos, Hoeft}.  Our result for $n=3$ is then a good
candidate to describe DMH. Indeed, the undesired feature of the cuspy behaviour led at least some 
groups to model the DMH as a polytropic configuration \cite{arieli, Dehnen, Matos, McKee, Debattista, Gonzalez, Yepes, Henriksen,
Hogan}. Sometimes it is claimed that a polytropic model is favoured over the results from N-body simulation \cite{Zavala}. However, the oscillatory behavior of the $n=3$ solutions of the $\Lambda$LE represents a disadvantage when compared with the NFW profiles, although the region of physical interest (below the virial radius) is well represented by the solutions  of  $\Lambda$LE.




\section{Conclusions}
In this paper we have explored the effects of a positive cosmological
constant on the equilibrium and stability of astrophysical
configurations with a polytropic equation of state.  We have found
that the radius of these kind of configurations is affected in the
sense that not all polytropic indices yield configurations with
definite radius even in the asymptotic sense. Among other, the widely
used isothermal sphere model becomes a non-viable model in the
presence of $\Lambda$ unless we are ready to introduce an arbitrary
cut-off which renders the model unappealing. Indeed, in this
particular case we have tried different definitions of a finite radius
with the result that none of them seems to be justified, either from
the phenomenological or from the theoretical point of view. This is
then an interesting global result: $\Lambda$ not only affects
quantitatively certain properties of large, low-density astrophysical
structures, but it also excludes certain commonly used models
regardless what density we use.

For polytropic indexes $n<5$ and for \emph{certain values} of the central density we
cannot find a well definite radius.  These \emph{certain values} are
encoded in a generalization of the equilibrium condition found for
spherical configurations (i.e, $\rho>2\rvac$) written now as
$\rho_{\rm c}>\maa_{n}\rvac$. We obtain $\maa_{1}=10.8$,
$\maa_{3/2}=24.2$, $\maa_{3}=307.7$, $\maa_{4}\approx 4000$.  These
values set a minimal central density for a given polytropic index
$n$. We have discussed such minimal density configurations and
determined their average density which strongly depend on $n$.
Interestingly, the radius of such configurations has a connection to
the length scale which appears in the Schwarzschild- de Sitter as the
maximally possible radius for bound orbits \cite{bala1}. In this
framework we found also a solution of very low, non-constant density.
Indeed, the limiting value of the central density in the above equations
is a crucial point. Below this value no matter can be in equilibrium. However,
above this value low density objects can still exist. Both effects
are due to $\Lambda$: in the first case the external repulsive force is
too strong for the matter to be in equilibrium, in the second case
this force can counterbalance the attractive Newtonian gravity effects (even if the
pressure is small).

Other examples of low density configuration which we examined in some
detail are  neutrino stars with mass of the order $1$ eV and $1$
keV.  In there we found that a nowadays dominating cosmological constant
affects both the mass as well as the radius of such exotic objects. 
Such effect could change those physical quantities several orders of magnitude.  
The magnitude of such effects, however, depend on the fermionic masses  and on the assumption that  
the fermions in such a configuration would be
essentially non-relativistic.

Finally, we made a conjecture regarding boson stars, and used variational methods in connection with the
Thomas-Fermi equations which could give relatively good results even without
invoking the whole general relativistic formalism,
We then found  extremely low density configurations for such astrophysical objects. 
Notice however, that this
conjecture relied purely on arguments based on scales and in fact needs a full
general relativistic investigation to confirm the results here obtained.

We have compared polytropic configurations with Dark Matter density profiles 
from N-body simulations. Surprisingly, we find a reasonable agreement between
both approaches for the polytropic index $n=3$ and restricting ourselves to the
virial radius. Our model does not have the the undesired features of the cuspy
behaviour of the NFW profiles.

The importance of the astrophysical properties and configurations
found in this article is that they are specific features to the
existence of a dark energy component.   Hence, such
configurations (e.g, low density configurations) or properties, if
ever found in nature would imply a strong evidence for the presence of
a dark energy component. Such observations would be a completely independent, and so complementary, 
of other cosmological probes of dark energy such as Supernova Ia or the CMBR.

\section*{Acknowledgments}
We acknowledge Stefanie Phleps for her comments on the manuscript. DFM acknowledge support from the A. Humboldt
Foundation.

\bibliographystyle{mn2e}

\begin{thebibliography}{}

\bibitem[Aldrovandi et al.1998]{aldro} Aldrovandi, R., Barbosa, A.,
L., Crispino, L., C., B. et al., Class. Quant. Grav. \textbf{16}
495-506 (1999)
\bibitem[Arieli 2003]{arieli}Arieli Y., Rephaeli Y., New Astronomy \textbf{8}, 517-528, 2003.
\bibitem[Balaguera-Antol{\'{\i}}nez \& Nowakowski 2005]{bala3}
Balaguera-Antol{\'{\i}}nez, A., Nowakowski, M., Astron. Astrophys
\textbf{441}, 23 (2005)
\bibitem[Balaguera-Antol\'{\i}nez \& Nowakowski 2006]{bala4}
Balaguera-Antol\'{\i}nez, A., Nowakowski, M., AIP Conf. Proc. {\bf 861}, 1001 (2006), arXive: astro-ph/0603624
\bibitem[Balaguera-Antol{\'{\i}}nez,B\"ohmer \& Nowakowski
2005a]{bala2} Balaguera-Antol{\'{\i}}nez, A.,B\"ohmer, C.,
Nowakowski, M., Int. J. Mod. Phys \textbf{D14}, 9, 1507-1526 (2005)
\bibitem[Balaguera-Antol{\'{\i}}nez,B\"ohmer \& Nowakowski
2006b]{bala1} Balaguera-Antol{\'{\i}}nez, A.,B\"ohmer, C.,
Nowakowski, M, Class. Quant. Grav \textbf{23}, 485-496 (2006)
\bibitem[Balaguera-Antol\'{\i}nez,Mota \& Nowakowski 2006]{bala5} Balaguera-Antol{\'{\i}}nez, A., Mota, D., F., Nowakowski, M.,
Class. Quant. Grav \textbf{23}, 4497-4510 (2006)
\bibitem[Baryshev,Chernin \& Teerikorpi 2001]{baryshev} Baryshev, Yu., Chernin, A., and Teerikorpi, P., Astron. Astrophys. {\bf 378},
729 (2001)
\bibitem[Binney \& Tremaine 1987]{binney} Binney, J $\&$ Tremaine, S.,  \emph{Galactic 
Dynamics}, Princeton University Press,  1987
\bibitem[Blake \& Glazebrook 2003]{blake}  Blake, C., and Glazebrook, K., Astrophys.\ J.\  {\bf 594}, 665 (2003)
\bibitem[B\"orner 2004]{boerner} B\"orner, G., \emph{The Early Universe}, 4th edition, Springer, 2004
\bibitem[B\"ohmer 2004]{boehmer1} B\"ohmer, C. G., Gen. Rel. Grav. \textbf{36}, 1039 (2004)
\bibitem[B\"ohmer \& Harko 2005]{boehmer}B\"ohmer, C. G. \& Harko T.,  Phys. Rev. \textbf{D71}, 084026 (2005), arXiv: astro-ph/0509874
\bibitem[ Brookfield et al. 2006]{brook} Brookfield, A.,  et al.  Phys.\ Rev.\ Lett.\  {\bf 96}, 061301 (2006)
\bibitem[Cabral-Roseti et al 2004]{sussman}Cabral-Roseti, L.G., Matos T., N\'u\~ez D., Sussman R., Zavala J., arXive:astro-ph/0405242.
\bibitem[Caimmi 2007]{caimmi} Caimmi R., arXive:gr-qc/0608030v1


\bibitem[Caldwell 2002]{cald} Caldwell, R. R., Phys.\ Lett.\ B {\bf 545}, 23 (2002)
\bibitem[Cardoso \& Gualtieri 2006]{cardoso}  Cardoso, V. $\&$ Gualtieri, L., Class.\ Quant.\ Grav.\  {\bf 23}, 7151 (2006)
\bibitem[ Chandrasekhar 1967]{chan} Chandrasekhar, S.  \emph{An Introduction to the Study of Stellar Structure}, Dover Publications, 1967.
\bibitem[Chavanis 2001]{chavanis1} Chavanis, P., Astron. Astrophys. \textbf{381}, 340 (2002), arXive: astro-ph/0103159

\bibitem[Chen \& Ratra 2004]{chen}   Chen G, and Ratra B, 2004 Astrophys.\ J.\  {\bf 612}, L1
\bibitem[Chernin Nagirner \& Starikova 2003]{chernin2} Chernin, A. D., Nagirner, D. I., Starikova, S. V., Astron. Astrophys. {\bf 399}, 19 (2003)
\bibitem[Chernin et al. 2007]{cher} Chernin {\it et al.}, arXiv:0704.2753 [astro-ph].
\bibitem[Daly \& Djorgovski 2004]{daly} Daly, R., A., and Djorgovski, S., G.,  Astrophys.J.  {\bf 612}, 652 (2004)
\bibitem[Daly \& Djorgovski 2003]{daly2}Daly, R., A., and Djorgovski, S., G., Astrophys. J.  {\bf 597}, 9 (2003)
\bibitem[Debnath et al 2006]{debnath} Debnath U., Nath, S., Chakraborty, S., Mon. Not. Roy. Astron. Soc., \textbf{369}, 1961 (2006)
\bibitem[Debattista \& Sellwood 1998]{Debattista} Debatissta, V.P. and Sellwood, J. A.,  Astropys. J. {\bf 493}, L5 (1998)
\bibitem[Dehnen \& Rose 1993]{Dehnen} Dehnen, H. and Rose, B., Astrophys. Space science {\bf 207}, 133 (1993)
\bibitem[Diemand et al 2007 ]{diemand} Diemand J., Kuhlen M., Madau P., submitted to Astron. Astrophys. ArXive: astro-ph/0703337
\bibitem[Dolgov \&  Hansen 2002]{dolgov} Dolgov A. D., Hansen S.H., Astropart.Phys. \textbf{16}, 339-344  (2002)
\bibitem[Eckehard \&  Schunck 1998]{eckehard} Eckehard W., Schunck F., arXiv: gr-qc/9801063
\bibitem[Einstein \& Straus 1945]{einstein} Einstein, A. Straus E. G., Rev. Mod. Phys \textbf{17}, 2 and 3, 1945
\bibitem[Gentile \& Tonini \& Salucci 2007]{gentile}Gentile G.F., Tononi C., Salucci P., Accepted in Astron. Astrophys
, arXive:astro-ph/0701550.
\bibitem[Gibbons \& Patricot 2003]{gibbons} Gibbons, G.W., Patricot C.E., Class.Quant.Grav. \textbf{20}, 5223 (2003)
\bibitem[Gonzalez-Casado et al. 2004]{Gonzalez} Gonzalez-Casado, G. at al. in Proceedings IAU Colloqium No. {\bf 195} 
(2004)   
\bibitem[Gruzinov 2000]{gruzinov} Gruzinov, A., Astrophys. J, \textbf{498}, 458 (1998), arXiv: astro-ph/9705026
\bibitem[Herrera \& Barreto 2003]{herrera} Herrera, L., Barreto, W., Gen.Rel.Grav. \textbf{36} 127-150 (2004), arXiv:gr-qc/0309052
\bibitem[Henriksen 2004]{Henriksen} Henriksen, R. N., Mon.Not.Roy.Astron.Soc. {\bf 355}, 1217 (2004)
\bibitem[Hoeft \& M\"uecket \$ Gottl\"ober 2004]{Hoeft} Hoeft, M>, M\"ucket, J. P. and Gottl|''ober, S., Astrophys. J.
{\bf 602}, 162 (2004)
\bibitem[Hogan \& Dalcanton 2000]{Hogan}Hogan, C. J. and Dalcanton, J. J., Phys. Rev. {\bf D62}, 063511 (2000)  
\bibitem[Horedt 2000]{Horedt} Horedt, G. P., Publ. Astron. Soc. Japan, \textbf{52}, 217 (2000)

\bibitem[Horellou \& Berge 2005]{hore}  Horellou  C, and Berge J., 2005, Mon.\ Not.\ Roy.\ Astron.\ Soc.\  {\bf 360}, 1393

\bibitem[Iorio 2005]{iorio} Iorio, L., Int.J.Mod.Phys. \textbf{D15} 473-476 (2006), arXiv:gr-qc/0511137
\bibitem[Jackson 1970]{jackson} Jackson J., 1970, Mon. Not. Roy. Astro. Soc. 148, 249 
\bibitem[Jetzer 1996]{jetzer} Jetzer P., astro-ph/9609068
\bibitem[Jetzer \& Serena 2006]{jetzer1} Jetzer, P. and Serena, M., Phys. Rev. {\bf D73}, 044015 (2006)
\bibitem[Kaniadakis,Lavagno \& Quarati 1996]{kaniadakis} Kaniadakis, C., Lavagno, A., Quarati, P., Phys.Lett. \textbf{B369} 308-312 (1996) arXiv: astro-ph/9603109 
\bibitem[Kagramanova,Kunz \& Laemmerzahl 2006]{kagramanova} Kagramanova, V., Kunz, J. and Laemmerzahl, C., Phys. Lett. {\bf B634}, 465 (2006)
\bibitem[Kawano et al. 2004]{kawano} Kawano, Y., Oguri, M., Matsubara, T., Ikeuchi, S., Publ.Astron.Soc.Jap. \textbf{56} 253-260 (2004)
\bibitem[Kennedy \&  Bludman 1999]{kennedy1}Kennedy, D.,  Bludman, S., Astrophys. J. \textbf{525} :1024-1031 (1999)
\bibitem[Koivisto \& Mota 2006]{koivisto}  Koivisto T., and  Mota D.F., Phys.\ Rev.\  D {\bf 73} (2006) 083502  [arXiv:astro-ph/0512135].
\bibitem[Koivisto \& Mota 2007a]{koivisto2} Koivisto T, \& Mota, D., F.,   Phys.\ Rev.\  D {\bf 75} (2007a) 023518 [arXiv:hep-th/0609155].
\bibitem[Koivisto \& Mota 2007]{Koivisto1}  Koivisto T., and  Mota D.F., Phys.\ Lett.\ B {\bf 644} (2007) 104
\bibitem[Koivisto \& Mota 2007b]{Koivisto3}  Koivisto T., and  Mota D.F., (2007b) arXiv:0707.0279 [astro-ph].
\bibitem[Lahav et al. 1991]{lahav91}Lahav O., et al.,Mon.Not.Roy.Astron.Soc \textbf{251}, 128-136 (1991)
\bibitem[Lai 2004]{chi} Lai, C., W,. arXiv: gr-qc/0410040
\bibitem[Lattanzi,Ruffini \&  Vereshchagin 2003]{lattanzi} Lattanzi M., Ruffini R.,  Vereshchagin G., AIP Conf.Proc. \textbf{668} (2003) 263-287
\bibitem[Lombardi \& Berti 2001]{lombardi}Lombardi, M., Bertin, G., Astron. Astrophys \textbf{375}, 1091-1099 (2001), arXiv: astro-ph/0106336
\bibitem[Lynden-Bell \& Wood 1968]{lynden} Lynden-Bell D., Wood R., Mon.Not.Roy.Astron.Soc. \textbf{138}, 495 (1968)
\bibitem[Maccio 2004]{maccio} Maccio A., Mon.Not.Roy.Astron.Soc. \textbf{361},  1250-1256 (2005), arXiv:astro-ph/0402657
\bibitem[Manera \& Mota 2005]{manera} Manera, M. and Mota, D. F., Mon.Not.Roy.Astron.Soc. \textbf{371} 1373 (2006), arXiv:astro-ph/0504519
\bibitem[Maor \& Lahav 2005]{ml} Maor I., Lahav O., arXiv:astro-ph/0505308
\bibitem[Matos et al. 2005]{Matos} Matos, T., Nunez, D. and Sussman, R., Gen. Relat. Gravit. {\bf 37}, 769 (2005)
\bibitem[Mota \& van de Bruck 2004]{mota1} Mota, D. F. and van de Bruck, C., Astron. Astrophys. {\bf 421}, 71 (2004)
\bibitem[Mota \& Shaw 2007]{shaw2} Mota D. F. and Shaw D. J.,  2007, Phys.\ Rev.\  D 75, 063501; 
\bibitem[Mota \& Shaw 2006]{shaw3} Mota D. F. and Shaw D. J.,  2006, Phys.Rev.Lett.97:151102
\bibitem[McKee 2001]{McKee} McKee, C. F. in ASP Conf. Proc. {\bf 243} (2001)
\bibitem[Mukhopadhyay \& Ray 2005]{mukhop}Mukhopadhyay, U., Ray S., arXiv: astro-ph/0510550
\bibitem[Natarajan \& Lynden-Bell 1997]{natarajan} Natarajan P., Lynden-Bell D., Mon.Not.Roy.Astron.Soc. \textbf{286}, 268-270 (1997)
\bibitem[Navarro \& Frenk \& White 1996]{nfw}Navarro, J.S., Frenk, C.S., White, S.D.M.,  Astrophys. J  \textbf{462}, 563 (1996) 
\bibitem[Noerdlinger \& Petrosian 1971]{noer}Noerdlinger P., Petrosian V., Astrophys. J,  \textbf{168}, 1 (1971)
\bibitem[Nojiri 2005]{nojiri}Nojiri S., Odinstov S. D., Phys.Rev. \textbf{D72} (2005) 023003, arXiv: hep-th/0505215
\bibitem[Nowakowski 2001]{nowakowski2} Nowakowski, M., Int. J. Mod. Phys., \textbf{D10}, 649 (2001)
\bibitem[Nowakowski \& Sanabria \& Garcia 2002]{nowakowski1} Nowakowski, M., Sanabria, J.-C., and Garcia, A., Phys. Rev. {\bf D66}, 023003 (2002)
\bibitem[Nunes \& Mota 2006]{nunes} Nunes, N. J. and Mota, D. F., Mon. Not. Roy. Astron. Soc. 368:2 751 (2006), arXive: astro-ph/0409481
\bibitem[Padmananbhan 1993]{padma} Padmananbhan, T., \emph{Structure Formation in the Universe}, Cambridge University Press, 1993
\bibitem[Penston 1969]{penston} Penston M.V.,  Mon.Not.Roy.Astron.Soc. \textbf{144}, 425 (1969)
\bibitem[Pinzon \& Calvo-Mozo 2001]{pinzon1} Pinzon, G., A.,  Calvo-Mozo, B., arXive: astro-ph/0107428 

\bibitem[Ren et al. 2006]{rem}Ren J., Li,\& Shen H.,arXive: astro-ph/0604227 

\bibitem[Riess et al. 2004]{riess} Riess, A. G. {\it et al.} [Supernova Search Team Collaboration],
Astrophys. J.  {\bf 607} 665 (2004)
\bibitem[Rines et al. 2002]{rines}Rines, K. et al.  Astron.J. \textbf{124} (2002) 1266, arXive: astro-ph/0206226
\bibitem[Ruffet et al. 1996]{ruffet} Ruffet, M., Rampp, M.,  Hanka, H.,Th., Astron. Astrophys \textbf{321}, 991-1006 (1997)
\bibitem[Ruffini \&  Bonazzola 1969]{ruffini} Ruffini R., Bonazzola S., Phys. Rev \textbf{187}, 5 (1969) 
\bibitem[Sadeth \& Rephaeli 2004]{sadeth} Sadeth, S., Rephaeli, Y., New Astron. \textbf{9}, 159-171 (2004) 
\bibitem[Seo \& Eisenstein 2003]{seo}  Seo, H., J., and Eisenstein, D., J., Astrophys.J.  {\bf 598}, 720 (2003)
\bibitem[Sereno 2005]{sereno} Sereno M., Mon.Not.Roy.Astron.Soc. \textbf{356}, 937-943 (2005)
\bibitem[Shapiro \& Teukolsky 1983]{shapiro} Shapiro, S. \& Teukolsky, S., \emph{Black Holes, White Dwarfs and Neutron Stars}, Wiley-Interscience Publications, 1983
\bibitem[Sommer-Larsen,Vedel \& Hellsten 1996]{sommer} Sommer-Larsen J., Vedel, H., Hellsten U., Astrophys.J. \textbf{500} (1998) 610-618
\bibitem[Spergel et al. 2006]{wmap3}Spergel D., {\it et al},  astro-ph/0603449
\bibitem[Spruch 1991]{spruch} Spruch L., Rev. Mod. Phys, \textbf{63}, 1 (1991) 
\bibitem[Shaw \& Mota 2007]{shaw} Shaw D. J. and Mota D. F., 2007, to appear in the Astrophys. J. Suppl. arXiv:0708.0868 [astro-ph]
\bibitem[Sussman \& Hernandez 2003]{more} Sussman R. A.
and Hernandez X.,  2003, Mon.Not.Roy.Astron.Soc. 345 871.
\bibitem[Tegmark et al. 2004]{tegmark}Tegmark M., {\it et al.}  [SDSS Collaboration], 
Astrophys.\ J.\  {\bf 606}, 702 (2004)
\bibitem[Umemura \& Ikeuchi 1986]{Umemura} Umemura, M. and Ikeuchi, S., Astron. Astrophys. \textbf{165}, 1 (1986)
\bibitem[Wang 2006]{wang3}Wang P. Astrophys. J.  {\bf 640}, 18-21 (2006)
\bibitem[Wang \& Mukherjee 2004]{wang2} Wang, Y., and Mukherjee P., Astrophys. J.  {\bf 606}, 654 (2004)
\bibitem[Wang \& Steinhardt 1998]{wang} Wang, L., and Steinhardt, P. J., Astrophys. J. {\bf 508}, 483 (1998)
\bibitem[Weinberg 1972]{weinberg}Weinberg, S., \emph{Gravitation and Cosmology}, Wiley and Sons, 1972
\bibitem[Yabushita 1968]{yabu} Yabushita S., Mon.Not.Roy.Astron.Soc. \textbf{140}, 109 (1968)
\bibitem[Yepes et al. 2004)]{Yepes}  Yepes, G. at al. in Proceedings IAU Colloquim {\bf 195} (2004)
\bibitem[Zavala et al. 2006]{Zavala} Zavala, J., Nunez, D., Sussman, R., Cabral-Rosetti, C. G. and Matos, T., 
J. Cosmol. Astrpart. Phys. {\bf 06}, 008 (2006)
\end{thebibliography}

\label{lastpage}
\end{document}